\documentclass[aps,prx,reprint,superscriptaddress]{revtex4-1}

\usepackage{scalerel}
\usepackage{amssymb}
\usepackage{suffix}
\usepackage{float}
\usepackage{mathtools}
\usepackage[utf8]{inputenc}
\usepackage{booktabs}
\usepackage{cases}
\usepackage[multiple]{footmisc}
\usepackage{dcolumn}
\usepackage{color,soul}
\usepackage{rotating}
\usepackage{perpage}
\usepackage{xcolor}
\usepackage{soul}
\usepackage{tikz}
\usepackage[T1]{fontenc}
\usepackage{etoolbox}
\usepackage{graphics}
\usepackage{siunitx}
\usepackage[hidelinks]{hyperref}
\hypersetup{
    colorlinks,
    linkcolor={red},
    citecolor={blue},
    urlcolor={blue}
}
\usepackage{float}	
\usepackage{collref}
\usepackage{multirow}
\usepackage{mathtools}
\usepackage{bm}
\usepackage{url}

\usepackage{tikz}
\usepackage{tikz-3dplot}
\usepackage{changes}

\newcommand{\red}[1]{\textcolor{black}{#1}}

\newcommand*\rfrac[2]{{}^{#1}\!/_{#2}}

\begin{document}
\title{Terahertz high-harmonic generation in gapped antiferromagnetic chains}

	\author{Mohsen Yarmohammadi}
	\email{mohsen.yarmohammadi@utdallas.edu}
 \affiliation{Department of Physics, The University of Texas at Dallas, Richardson, Texas 75080, USA}
 \author{Michael H. Kolodrubetz}
	\email{mkolodru@utdallas.edu}
	\affiliation{Department of Physics, The University of Texas at Dallas, Richardson, Texas 75080, USA}
	
	\date{\today}
	
	\begin{abstract}
The nonlinear dynamics of magnetization in antiferromagnets, resulting in high-frequency spin waves~(high-order harmonics) as signal carriers, enable fast magnetic state switching in spintronic devices. More harmonic orders potentially allow more information to be conveyed by the spins. Developing theoretical models to describe these waves in antiferromagnets is essential for predicting their properties and guiding experimental efforts. Here, we consider the role of linear and quadratic spin-phonon couplings (SPCs) in achieving high-order harmonics in the THz magnetization of a gapped antiferromagnetic spin chain. A THz steady laser's electric field indirectly drives spins via phonons. Using spin-wave theory, mean-field theory, and the Lindblad formalism, we analyze the resulting nonlinear dynamics. We highlight the distinct mechanisms for harmonic generation when a phonon is coupled to the easy-plane and easy-axis of spins. Moreover, we observe that quadratic SPC blocks odd harmonics due to invariant inversion symmetry, while linear SPC generates both odd and even harmonics. We also investigate the effects of drive frequency, drive amplitude, phonon damping, and spin damping on the number of harmonics. Our findings offer an alternative pathway for developing nonlinear magnonics.
	\end{abstract}
	\maketitle
	{\allowdisplaybreaks
 
		\section{Introduction}
  
  The advantages of spintronic systems over electronic counterparts are well-established across various science and technology domains, including quantum computing and information processing~\cite{RevModPhys.76.323,HIROHATA2020166711,YUAN20221}. Antiferromagnetic spin systems hold significant promise for advancing applications, owing to their enhanced stability and reduced energy consumption for rapid spin manipulation~\cite{magnetochemistry8040037,Chumak2015,Jungwirth2016,RevModPhys.90.015005,uhrig2024landaulifshitz,PRXQuantum.4.030332,PhysRevB.109.174419,PhysRevB.104.224424,Bossini_2020,PhysRevB.104.184419}. In the realm of antiferromagnets, the investigation of high-harmonic generation (HHG) has attracted considerable attention, offering a potent tool for probing and manipulating spin dynamics on ultrafast timescales~\cite{10.1063/1.2199473,doi:10.1126/science.aab1031,Nemec2018,Kampfrath2011,doi:10.1126/sciadv.aar3566,Li2020,doi:10.1126/science.aaz4247,10.1063/5.0075999,Mrudul2020,PhysRevB.109.144418,PhysRevB.108.064427}. Recent progress in ultrafast spectroscopy and time-resolved techniques has provided valuable insights into the intricate interplay between spin, lattice~(phonon), and electronic properties in antiferromagnetic materials. However, effectively probing the nonlinear spin responses in antiferromagnetic materials using optical fields has posed challenges, mainly due to their small magneto-optical susceptibility~\cite{Zhang2023,Schlauderer2019}.
  
   Spin-phonon coupling~(SPC) offers a notable route to achieve novel magnetic phases and manipulate spin configurations in magnetic materials~\cite{hart2024phonondriven,PhysRevLett.100.077201,Hernandez_2023,PhysRevB.94.014409}. In traditional magnets, localized Einstein phonons can significantly influence the spin-exchange interaction between bonds~\cite{PhysRevResearch.4.013129,PhysRevResearch.5.013204,PhysRevB.109.224417,allafi2024spin,strungaru2023route,zhou2024coherent}. Recent progress in pump-probe experiments and optics allows for resonantly exciting various phonon modes and investigating their effects on both electron and spin dynamics~\cite{PhysRevB.97.094108,Ginsberg2023,PhysRevB.106.064303,zheng2024phononmediated,PhysRevA.106.053116}. Although numerous studies have explored the potential of spin systems in higher harmonic generation~\cite{zhang2018generating,PhysRevB.102.081121,PhysRevB.99.184303,kanega2021linear,allafi2024spin,ZHAO2021449}, addressing on-demand HHG in antiferromagnets continues to raise several intriguing questions that researchers are actively pursuing.

In this paper, we pose the question: how effectively do the linear and nonlinear SPCs, along with the easy-plane and the easy-axis of spin interactions, generate higher harmonics in gapped antiferromagnetic chains with inversion symmetry? To address this, in contrast to previous research with intense THz magnetic fields~\cite{Zhang2023,Mukai_2016,STREMOUKHOV2024107377,Mukai_15,PhysRevB.106.014420}, \red{we drive the system by a weak THz electric field. We assume an ideal case where the laser is perfectly periodic. 
Maintaining inversion symmetry in the chain excludes a linear coupling between the laser and the Heisenberg spin-spin interaction $\vec{S} \cdot \vec{S}$. However, it does not exclude a linear coupling between phonon displacement and $\vec{S} \cdot \vec{S}$ when the atomic motion has a component perpendicular to the spin-bond (our assumption for a general geometry). Thus, the mechanism focuses on generating high-order harmonics of the pump frequency through indirect phonon-induced spin excitations~\cite{Shih2013,Hlubek_2012,PhysRevLett.116.017204,PhysRevB.92.054409}. In doing so, we employ spin-wave theory, mean-field theory, and a Lindblad formalism to treat the time evolution of both spin and phonon sectors in the presence of a quantum dissipation~\cite{breuer2007theory,lindblad1976}.} Our focus veers away from long-range magnetic ordering, as it typically lacks stability in low-dimensional magnets. Therefore, to make progress with a straightforward perturbative treatment of the spin system, we use a chain with a substantial anisotropy such as a Sm-Fe chain in the crystal SmFeO$_3$~\cite{PhysRevB.109.224417} having both linear and nonlinear SPCs. Our model maintains sufficient generality for broader materials in realms of cold atom physics, which has made the realization of higher-spin quantum systems with strong anisotropy increasingly plausible~\cite{Altman_2003,PhysRevLett.126.163203,PhysRevLett.128.093401}. 

This paper is organized as follows. Section~\ref{s2}, first, outlines the model of a gapped antiferromagnetic spin chain. Then, it calculates non-equilibrium magnetization. In Sec.~\ref{s3}, we present the main findings obtained for different parameters, \red{taking into account a continuous wave laser.} We also briefly discuss the specific measurement of ultrafast magnetization in the experiment. Finally, Sec.~\ref{s4} concludes the key results of the paper.
    
		\section{Model and methods}\label{s2} \begin{figure}[t]
		 	\centering
		 	\includegraphics[width=1\linewidth]{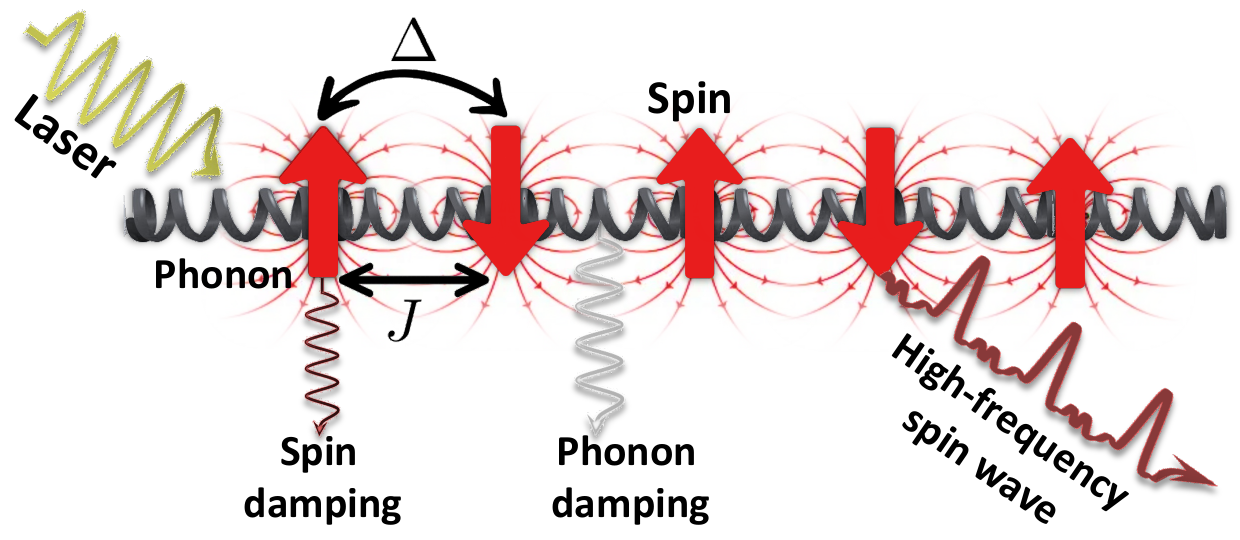}
        \caption{\textbf{Schematic of a driven-dissipative gapped antiferromagnetic spin chain.} A chain of spins, alternating in direction, indicates antiferromagnetic ordering with easy-plane~($J$) and the easy-axis   ~($\Delta$) interactions. Black springs connecting the spins represent optical phonons. Damping mechanisms, depicted as light vertical waves, connect the spins and phonons to the environment (bath). A steady laser field illuminates the chain, driving the phonons, and inducing higher harmonics in the magnetization dynamics through linear and quadratic spin-phonon couplings, see the text for details.} 
		 	\label{f1}
		 \end{figure}
		In this section, we first depict a schematic of a driven-dissipative gapped antiferromagnetic spin chain comprising phonon and spin sectors in Fig.~\ref{f1} and then model the individual chain's components through approximations. Next, we calculate the magnetization dynamics.

  \subsection{Time-dependent Hamiltonian}
  \subsubsection{Phonon and phonon-laser coupling}
        In the phonon sector, we consider optical phonons represented by bosonic creation and annihilation operators $a^\dag$ and $a$, originating from lattice displacements. We assume they are non-interacting zero-momentum infrared-active optical phonons, that are described by
        \begin{equation}
            \mathcal{H}_{\rm p} = {} \omega_{\rm p} a^\dagger a\, ,
        \end{equation}with energy $\hbar \omega_{\rm p}$~(we set $\hbar = 1$)~\cite{yarmohammadi2024ultrafast,yarmohammadi2020dynamical,PhysRevB.109.224417,yarmohammadi2023nonequilibrium,PhysRevB.107.174415}, which has the highest phonon energy at small momenta compared to dispersive ones. 

         In the phonon-laser coupling sector, we introduce the optical phonons stimulated by the THz laser's electric field, described by~\cite{PhysRevB.95.205111,PhysRevB.104.L220503,Kennes2017,yarmohammadi2024ultrafast,yarmohammadi2020dynamical,PhysRevB.109.224417,yarmohammadi2023nonequilibrium,allafi2024spin,PhysRevB.107.174415,RevModPhys.93.041002,murakami2023photoinduced}
  \begin{equation}
      \mathcal{H}_{\rm lp}(t) =  {} E_0 \cos(\omega_{\rm d}\,t) \sqrt{N} (a^\dagger + a)\, ,
  \end{equation}where a steady laser drive is employed with amplitude $E_0$ and frequency $\omega_{\rm d}$ for the chain length of $N = 2001$. Specifically, we assume that the laser is polarized along the direction of nuclear motion, such that the electric field maximally couples to phonon displacement. We are primarily interested in the case where such a steady state laser is quenched on time $t=0$, for which we determine HHG in the non-equilibrium steady state (NESS). 

  \subsubsection{Spin and spin-phonon coupling}
  In the spin sector, we model an anisotropic antiferromagnetic spin $S > 1/2$ chain with Heisenberg spin-spin interactions between two sublattices $\ell$ and $j$, that is given by
		\begin{equation}\label{eq_1}
			\mathcal{H}_{\rm s} = {} J\sum_{\langle \ell,j \rangle} \bigg[ \frac{1}{2} \left(S^+_\ell S^-_j + S^-_\ell S^+_j \right)+\Delta S^z_\ell S^z_j\bigg]\, ,
		\end{equation}where $S^{\pm } = S^x \pm i S^y$. Here, $J$ on the order of THz represents the in-plane easy-plane exchange coupling, and $\Delta > 1$ (easy-axis) is the phenomenological the easy-plane parameter. This formulation considers nearest-neighbor interactions and approximates spins as classical vectors, with axes in spin space rotated for sites on the down sublattice to align with the classical N\'eel state~\cite{PhysRevB.109.224417}. We consider the system at very low temperatures, with a low population of excited magnons due to SPC, and employed many sites $N = 2001$~(to eliminate finite-size effects) with large spin lengths $S>1/2$ for both magnetic sublattices. In this case, the zeroth-order expansion of the Holstein-Primakoff transformation~\cite{PhysRev.58.1098} is suited for the linear spin-wave theory, to analyze spin excitations. Thus, we adopt the transformations $S^z_\ell = b^\dagger_\ell b_\ell - S$, $S^+_\ell \simeq \sqrt{2S}\,b_\ell$, $S^-_\ell \simeq  \sqrt{2S} \,b^\dagger_\ell$, 
				$S^z_j = -b^\dagger_j b_j + S$, $S^+_j \simeq \sqrt{2S}\,b^\dagger_j$, and $S^-_j \simeq \sqrt{2S} \,b_j$, where $b$ is the bosonic operator at two different sublattices $\ell$ and $j$. To derive the magnon dispersion, we diagonalize the spin Hamiltonian using the bosonic Bogoliubov transformation $b_k = {} \cosh(\theta_k) \tilde{b}_k +  \sinh(\theta_k) \tilde{b}^\dagger_{-k}$ with $\sinh(2 \theta_k) = - \cos(k)/\sqrt{\Delta^2-\cos^2(k)}$. Consequently, we find $\mathcal{H}_{\rm s}  = {} \sum_k \omega_k \tilde{b}^\dagger_k \tilde{b}_k + \mathcal{E}_0$, where $\omega_k = {} 2 J S\sqrt{\Delta^2 - \cos^2(k) }$ represents the magnon dispersion and $\mathcal{E}_0 = {} \frac{1}{2} \sum_k \omega_k - N J \Delta (S +1)$ denotes the ground state energy of magnons. 
  
  In the SPC sector, we aim to locally couple spins to phonons on each site. While the phonon, serving as the mediator between the driving laser and the spin system, can potentially couple with magnetic interactions through various mechanisms, we consider the Taylor expansion of photon-displacement-dependent exchange couplings to obtain linear and quadratic SPC. This choice represents an approximation of the real material. This coupling is driven by the relative oscillations of sublattices, influencing both easy-plane ($J$) and easy-axis ($\Delta$) interactions~\cite{PhysRevB.109.224417}. Hence, we employ $\mathcal{H}^{J}_{\rm sp} = \sum_{\langle \ell,j \rangle} \sum_{\alpha = x,y} {\big[g_{\rm l} (a^\dagger +a)+}g_{\rm q} (a^\dagger +a)^2{\big]}S^\alpha_\ell  S^\alpha_j$ and $\mathcal{H}^{\Delta}_{\rm sp} = \sum_{\langle \ell,j \rangle} {\big[g_{\rm l}\Delta (a^\dagger +a)+}g_{\rm q}\Delta (a^\dagger +a)^2{\big]}S^z_\ell  S^z_j$, where $g_{\rm l}$~($g_{\rm q}$) represents the linear (quadratic) SPC strength. Notably, both SPC and spin-spin interaction terms are even under inversion symmetry, allowing such coupling as a second-order spin effect for nonlinear spin dynamics. In various magnets, nonlinear spin dynamics originate from diverse sources. For instance, in a recent study, we showed that linear coupling of an optical phonon to intradimer and interdimer magnetic couplings generates higher harmonics in a dimerized spin $S = 1/2$ chain~\cite{allafi2024spin}. In contrast, this work highlights the role of both linear and quadratic SPCs, as well as easy-plane and easy-axis interactions, in producing higher harmonics in a gapped antiferromagnetic spin $S > 1/2$ chain.  Using a mean-field approximation to decouple spin and phonon excitations~\footnote{The quantum fluctuations proportional to $1/\sqrt{N}$ become negligible in a long chain $N =2001$, given that the phonon number scales with $N$}, we obtain
    \begin{equation}
    \begin{aligned}
				\mathcal{H}^{\rm MF}_{\rm sp}= {} & \bigg[\frac{g_{\rm l}}{\sqrt{N}}\langle a^\dagger+ a \big \rangle + \frac{g_{\rm q}}{N}\big \langle  (a^\dagger+ a)^2 \big \rangle {\bigg]} \times \\ {} &\times \sum_k  \Big( \mathcal{E}_k  +A_k \tilde{b}^\dagger_k\tilde{b}_k +B_k  \big[\tilde{b}_k \tilde{b}_{-k}+ \tilde{b}^\dagger_{-k}\tilde{b}^\dagger_k \big] \Big)\, ,
        \end{aligned}
		\end{equation}where $A_k = \{-4 J S \, \cos^2(k) /\omega_k,4 J S\, \Delta^2 /\omega_k\}$ and $B_k = \{2 J S \Delta \cos(k)/\omega_k,-2 J S \Delta   \cos(k) /\omega_k\}$ stand for \{easy-plane $J$, the easy-axis $\Delta$\} interactions. In the above Hamiltonian, $\mathcal{E}_k$ only describes the ground energy. 
  
    Here, suppose the easy-plane and easy-axis interactions are combined with equal magnitude. In that case, it will result in no spin waves due to the vanishing $B_k$ matrix element, while maintaining a finite local spin density. \red{Thereby, in this SPC formulation, we separately consider the coupling of phonons to easy-plane and easy-axis interactions, without explicitly considering their combined roles in generating SPC. This model is similar to our previous work on SmFeO$_3$~\cite{PhysRevB.109.224417}.}

\subsection{Nonlinear dynamics of magnetization and high-harmonic generation}
        Next, we calculate the dynamics of sublattice magnetization per spin by calculating the spin dynamics of each sublattice, given by $\mathcal{M}(t) = \frac{1}{N} \sum_{\langle \ell,j\rangle} \langle S^z_\ell\rangle(t) - \langle S^z_j\rangle(t) = \mathcal{M}_0 + \delta \mathcal{M}(t)$, where $\mathcal{M}_0 = S +\frac{1}{2}- \frac{2 J \Delta}{N} \sum_k \frac{1}{\omega_k}$ and
		\begin{equation}\label{eq_3}
            \delta\mathcal{M}(t) = \frac{J S }{2 N} \sum_k \left[\frac{\cos(k) }{\omega_k} {\rm Re}\langle  \tilde{b}^\dagger_{-k}  \tilde{b}^\dagger_k \rangle (t)-\frac{2 \Delta}{\omega_k} \langle  \tilde{b}^\dagger_{k}  \tilde{b}_k \rangle (t) \right]\, ,
		\end{equation}where the time evolutions of $\langle  \tilde{b}^\dagger_{-k}  \tilde{b}^\dagger_k \rangle (t)$ and $\langle  \tilde{b}^\dagger_{k}  \tilde{b}_k \rangle (t)$ are calculated via the Lindblad quantum master equation~\cite{breuer2007theory,lindblad1976}: $\dot{O} (t) =  {} i[\mathcal{H}(t),O]  + \sum_{j} \gamma_{j} \Big[{L}_{j}^{\dagger}O {L}_{j}  -\frac{1}{2}\big\{{L}_{j}^{\dagger}{L}_{j},O\big\}\Big](t)$, where $\mathcal{H}(t) = \mathcal{H}_{\rm p} + \mathcal{H}_{\rm lp}(t) + \mathcal{H}_{\rm s} + \mathcal{H}^{\rm MF}_{\rm sp}$. In this formalism, ${L}_{j} = \{a,\tilde{b}_k\}$ represents the set of time-independent Lindblad jump operators for both phonon and spin damping, see Fig.~\ref{f1}, with the decay rates of, respectively, $\gamma_{j} = \{\gamma_{\rm p},\gamma_{\rm s}\}$ at zero temperature. 
  
  Let us derive the time evolution of observables ($O$), accounting for the coupling of phonons to both the easy-plane and the easy-axis magnetic interactions: \begin{widetext}
		\begin{subequations} \label{eq_13} 
	\begin{align}  
		&\hspace*{0cm}\frac{d}{dt} \langle a^\dagger  + a \rangle (t) =  + i\omega_{\rm p} \, \langle a^\dagger  - a \rangle (t) -  \frac{\gamma_{\rm p}}{2} \langle a^\dagger  + a \rangle (t)\, ,\label{eq_8a}\\ 
		&\hspace*{0cm}\frac{d}{dt} \langle a^\dagger  - a \rangle (t) = +i \omega_{\rm p} \langle a^\dagger  + a \rangle (t)+ 4 i g_{\rm q}\widetilde{S}(t)\,\langle a^\dagger  + a \rangle (t)+2i\sqrt{N}\left(E(t)+ g_{\rm l}\widetilde{S}(t)\right)-  \frac{\gamma_{\rm p}}{2} \langle a^\dagger  - a \rangle (t),\label{eq_8b}\\
		&\hspace*{0cm}\frac{d}{dt} \langle a^\dagger a \rangle (t) =  -i \sqrt{N}\left(E(t)+  g_{\rm l}\widetilde{S}(t)\right)\, \langle a^\dagger  - a \rangle (t) - 2i g_{\rm q} \widetilde{S}(t)\langle a^\dagger a^\dagger  - aa \rangle (t) - \gamma_{\rm p} \langle a^\dagger a \rangle (t) \, ,\label{eq_8c}\\
		&\hspace*{0cm}\frac{d}{dt} \langle a^\dagger a^\dagger  + aa \rangle (t) = + 2 i \omega_{\rm p} \langle a^\dagger a^\dagger  - aa \rangle (t)+ 4 i g_{\rm q} \widetilde{S}(t)\,\langle a^\dagger a^\dagger - aa \rangle (t) + 2i \sqrt{N}\left(E(t)+  g_{\rm l}\widetilde{S}(t)\right)\, \langle a^\dagger  - a \rangle (t) \notag \\ {} & \hspace{2.95cm}- \gamma_{\rm p} \langle a^\dagger a^\dagger + aa \rangle (t)\, ,\label{eq_8d}\\
  &\hspace*{0cm}\frac{d}{dt} \langle a^\dagger a^\dagger  - aa \rangle (t) = + 2 i \omega_{\rm p} \langle a^\dagger a^\dagger + aa \rangle (t)+ 4 i \frac{g_{\rm q} }{N}\widetilde{S}(t)\,\langle a^\dagger a^\dagger + aa \rangle (t) + 2i \left(E(t)\sqrt{N}+  \frac{g_{\rm l}}{\sqrt{N}}\widetilde{S}(t)\right)\, \langle a^\dagger  + a \rangle (t) \notag \\ {} & \hspace{2.95cm}- \gamma_{\rm p} \langle a^\dagger a^\dagger - aa \rangle (t)\, ,\label{eq_8e}\\
		&\hspace*{0cm}\frac{d}{dt} \langle \tilde{b}^\dagger_k  \tilde{b}_k\rangle(t) = {} +2\left[\frac{g_{\rm l}}{\sqrt{N}}\langle a^\dagger +a \rangle (t)+ \frac{g_{\rm q}}{N}   \langle (a^\dagger +a)^2 \rangle (t) \right]  B_k {\rm Im}\langle \tilde{b}^\dagger_{-k}  \tilde{b}^\dagger_{k} \rangle (t)-\gamma_{\rm s} \langle \tilde{b}^\dagger_k  \tilde{b}_k\rangle(t)\, ,\label{eq_a1f}\\
		&\hspace*{0cm}\frac{d}{dt} \langle \tilde{b}^\dagger_{-k}  \tilde{b}^\dagger_{k} \rangle (t) = +2i\left(\omega_k +  \left[\frac{g_{\rm l}}{\sqrt{N}}\langle a^\dagger +a \rangle (t)+ \frac{g_{\rm q}}{N}   \langle (a^\dagger +a)^2 \rangle (t) \right]  A_k\right)\, \langle \tilde{b}^\dagger_{-k}  \tilde{b}^\dagger_{k} \rangle (t) \notag \\ {} &\hspace{2.35cm}+ 2i \left[\frac{g_{\rm l}}{\sqrt{N}}\langle a^\dagger +a \rangle (t)+ \frac{g_{\rm q}}{N}   \langle (a^\dagger +a)^2 \rangle (t) \right]   \left[\langle \tilde{b}^\dagger_{k}  \tilde{b}_k \rangle (t)+\frac{N}{2}\right]B_k - \gamma_{\rm s}\langle \tilde{b}^\dagger_{-k}  \tilde{b}^\dagger_{k} \rangle (t)\, ,\label{eq_8f}
	\end{align} 
\end{subequations} \end{widetext}The above equations are, respectively, for the phonon displacement $\langle a^\dagger  + a \rangle/\sqrt{N}$, phonon momentum $i\langle a^\dagger  - a \rangle/\sqrt{N}$, phonon number $\langle a^\dagger a \rangle/N$, squeezed phonon displacement $\langle a^\dagger a^\dagger  + a a \rangle/N$, squeezed phonon momentum $i\langle a^\dagger a^\dagger  - a a \rangle/N$, $k$-component of the spin density $\langle \tilde{b}^\dagger_{k}  \tilde{b}_{k} \rangle/N$, and the $k$-component of off-diagonal spin excitations (pair magnons) $\langle \tilde{b}^\dagger_{-k}  \tilde{b}^\dagger_{k} \rangle/N$, where \begin{equation}
        \widetilde{S}(t) = {} \frac{1}{N}\sum_k \left[A_k \langle \tilde{b}^\dagger_k  \tilde{b}_k \rangle (t)+ B_k {\rm Re} \langle \tilde{b}^\dagger_{-k}  \tilde{b}^\dagger_{k} \rangle (t)\right]\, .
\end{equation}
  
        We solve the coupled Eqs.~\eqref{eq_8a}-\eqref{eq_8f} numerically for various parameter sets and then plug in the solutions of spin excitations, $\langle \tilde{b}^\dagger_{-k} \tilde{b}^\dagger_k \rangle (t)$, and spin density, $\langle \tilde{b}^\dagger_{k} \tilde{b}_k \rangle (t)$, to Eq.~\eqref{eq_3} to compute the magnetization dynamics. The solutions strongly depend on the interaction-dependent matrix elements $A_k$ and $B_k$ along with both SPCs. 
        
        To investigate the complex superposition of frequencies~(i.e., HHG) in the NESS of magnetization dynamics, we employ the Fourier transform. We opt to subtract $\mathcal{M}_0$ from the total response for simplicity, allowing us to focus on the temporal treatment of the magnetization. Accordingly, the NESS driven by any frequency $\omega$ in $\delta \mathcal{M}(t)$ can be represented by the following Fourier series:\begin{equation}
\label{eq_4}
\delta \mathcal{M}(t) = {} \sum_m \delta \mathcal{M}_m e^{i m \omega t}\, ,
\end{equation}where $m \neq 0$ is the harmonic number. 

\section{Results and discussion}\label{s3}
To produce the strongest nonlinear responses in our model, we primarily consider resonant laser and phonon frequencies, setting $\omega_{\rm d} = \omega_{\rm p}$; see Appendix~\ref{apa} for weaker off-resonance $\omega_{\rm d} \neq \omega_{\rm p}$ excitations. Furthermore, we select parameters in units of $J$~(on the order of THz) ensuring the achievement of a NESS and the highest number of harmonics. These parameters are chosen to create a finely balanced simulation where exchange interactions, coupling effects, and damping all play significant roles in generating harmonics, but still combined effects of linear and quadratic SPCs overwhelmingly dominate to induce nonlinear effects and substantial HHG spectrum.
   
   Figure~\ref{f2} shows the ultrafast~(e.g., for $J = 1$ THz, NESS timescale is $t \approx 47$ picoseconds) dynamics of magnetization in an antiferromagnetic spin $S > 1/2$ chain, displaying different aspects of the easy-plane and easy-axis interactions as magnetization strongly depends on $A_k$ and $B_k$ matrix elements. We observe a series of short- and long-time dynamics oscillations. Since the easy-axis (anisotropy) leads to highly directional spin waves, phonon coupled to the easy-axis spins exhibits more complex NESS signals than the easy-plane case. The presence of high-harmonic components in the long-time dynamics of magnetization is evident, suggesting that the system's response to laser excitation involves complex, nonlinear processes beyond simple linear responses. 
   \begin{figure}[t]
		 	\centering
		 	\includegraphics[width=0.95\linewidth]{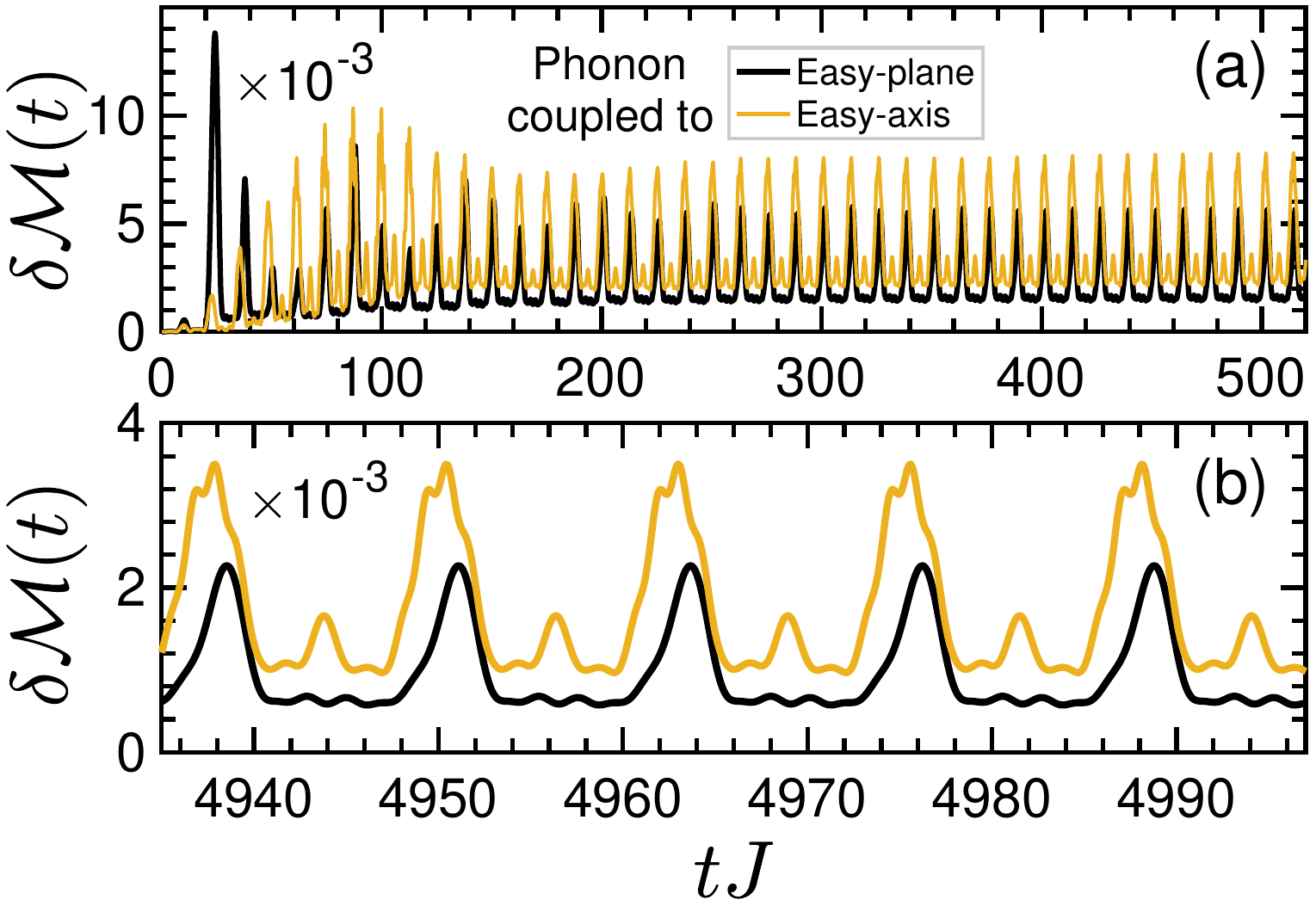}
        \caption{\textbf{The nonlinear dynamics of magnetization in a gapped antiferromagnetic spin chain.} (a) Short-time dynamics capture the immediate response of magnetization following laser excitation. (b) Long-time dynamics showcase the evolution of magnetization over an extended duration, revealing the presence of high-harmonic components or complex signals in the NESS. The parameters used in the simulations are $\omega_{\rm d} = \omega_{\rm p} = J/2$, $g_{\rm l} = g_{\rm q} = J/4$, $\gamma_{\rm p} = J/40$, $\gamma_{\rm s} = J/100$, and $E_0 = J/50$.} 
		 	\label{f2}
		 \end{figure}
   
   In general, from the time evolution of spin density $\langle \tilde{b}^\dagger_{k}\tilde{b}_{k} \rangle$ in Eq.~\eqref{eq_a1f}, it is apparent that the cross-talk between phonon displacement $\langle a^\dagger + a\rangle$, squared phonon displacement $\langle (a^\dagger + a)^2\rangle$, and spin excitation $\langle \tilde{b}^\dagger_{-k}\tilde{b}^\dagger_{k} \rangle$ introduces nonlinearity and HHG to the magnetization. This nonlinearity appears differently for various contributions, including those from linear and quadratic SPCs, as well as the easy-plane and the easy-axis interactions due to $A_k$ and $B_k$ matrix elements. The cross-talk between phonon displacement~(with dominant first harmonic $q_1$) and magnon pairs in the NESS can be approximately described by the following terms \begin{subequations}
       \begin{align}
           \langle a^\dagger + a\rangle(t) \approx {} & q_1 \cos(\omega_{\rm d} t)\, ,\\
           \langle \tilde{b}^\dagger_{-k}\tilde{b}^\dagger_{k} \rangle(t) \approx {} & \sum_{m \neq 0} C_m \cos([2m-1] \omega_{\rm d} t)\, ,
       \end{align}
   \end{subequations}where the $C_m$ coefficients are parameter-dependent Bessel functions of different kinds~\cite{allafi2024spin}, which are responsible for generating both odd and even higher harmonics in the equations of motion. This has been previously corroborated both numerically and analytically in a study conducted by some of us~\cite{allafi2024spin}. To avoid repetition, we omit the details here. A similar scenario can be applied to the cross-talk between $\langle (a^\dagger + a)^2\rangle$ and $\langle \tilde{b}^\dagger_{-k}\tilde{b}^\dagger_{k} \rangle$, resulting in only even harmonics.\begin{figure}[t]
		 	\centering
		 	\includegraphics[width=1\linewidth]{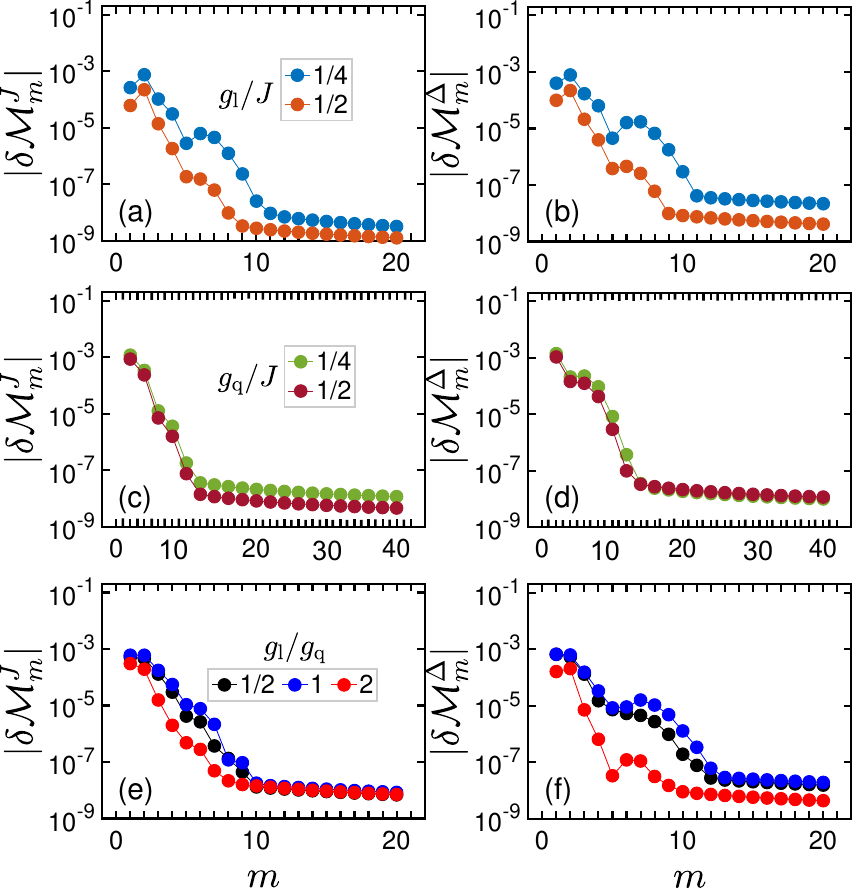}
        \caption{\textbf{Spin-phonon coupling effect on the HHG spectrum in nonlinear magnetization.} HHG spectrum for various SPCs for \{(a),(c),(e)\} easy-plane ($J$) and \{(b),(d),(f)\} the easy-axis ($\Delta$) interactions in the NESS. Increasing the SPCs reduces the number of harmonics, with a stronger effect observed for the easy-axis interaction. Linear~(quadratic) SPC leads to both odd and even (only even) harmonics due to the inversion symmetrical nature of couplings. Both linear and quadratic SPCs are essential for generating more harmonics. The parameters are fixed at $\omega_{\rm d} = \omega_{\rm p} = J/2$, $\gamma_{\rm p} = J/20$, $\gamma_{\rm s} = J/100$, and $E_0 = J/100$.} 
		 	\label{f3}
		 \end{figure}

   Let us extract the Fourier components for the HHG in the late-time dynamics of nonlinear magnetization across different parameter sets. While $\delta \mathcal{M}(t)$ in Eq.~\eqref{eq_4} with an integer subscript typically denotes a Fourier component with real and imaginary parts, we plot the absolute value of all these components for clarity in the following analysis. 
   
   We start with Fig.~\ref{f3} for various SPCs, demonstrating how increasing SPCs influence the harmonic generation in a gapped antiferromagnetic spin chain. \red{We briefly note that for strong SPCs, we need a weaker laser electric field to establish a NESS, caused by inherent instability stemming from the competition between the drive strength and the other set of parameters in the equations of motion, which has been verified numerically and can be verified analytically following the same procedure done by some of us before~\cite{yarmohammadi2020dynamical,PhysRevB.107.174415,allafi2024spin}}. Both interactions display a similar series of peaks representing harmonics, with a more pronounced impact observed for the easy-axis interaction. Since the laser field is considered an external time-dependent electric field that drives the usual matter-only constituent (lattice), we anticipate the intensity of generated harmonics to decrease for a fixed pump field.
 
The most commonly used methods for adjusting SPCs include temperature variations, doping, pressure alterations, and electrostatic gating~\cite{Mori2019,PhysRevMaterials.4.104802,Singh2019}. Since precise values of SPCs remain uncertain, we sweep $g_{\rm l}$ and $g_{\rm q}$ to identify overarching phenomena. \red{As the linear SPC~($g_{\rm l}$) increases, there is a reduction in the number of harmonics generated, see Figs.~\ref{f3}(a) and~\ref{f3}(b), since a strong $g_{\rm l}$ leads to nonlinear behaviors beyond perturbation theory. This reverses the monotonic dependence of HHG strength on SPC strength, as observed in other spin models as well~\cite{yarmohammadi2020dynamical,PhysRevB.107.174415}.} As the quadratic SPC~($g_{\rm q}$) increases, the change is negligible such that both bonds remain unaffected since $g_{\rm q}$ does not affect the inversion symmetry, see Figs.~\ref{f3}(c) and~\ref{f3}(d). However, as both linear and quadratic SPCs change, there is a noticeable change in both intensity and number of harmonics, see Figs.~\ref{f3}(e) and~\ref{f3}(f), meaning that the presence of both linear and quadratic SPCs is necessary to generate more harmonics compared to either one alone. However, this is invalid for $g_{\rm l}/g_{\rm q} \ll 1$ and $g_{\rm l}/g_{\rm q} \gg 1$. This, in turn, means that the linear SPC term dominates the response at weak driving fields, leading to strongly nonlinear effects. Above a critical harmonic, the system reaches a saturation point where the magnetization does not produce more harmonics and this differs in various parameters. \begin{figure}[b]
		 	\centering
		 	\includegraphics[width=0.95\linewidth]{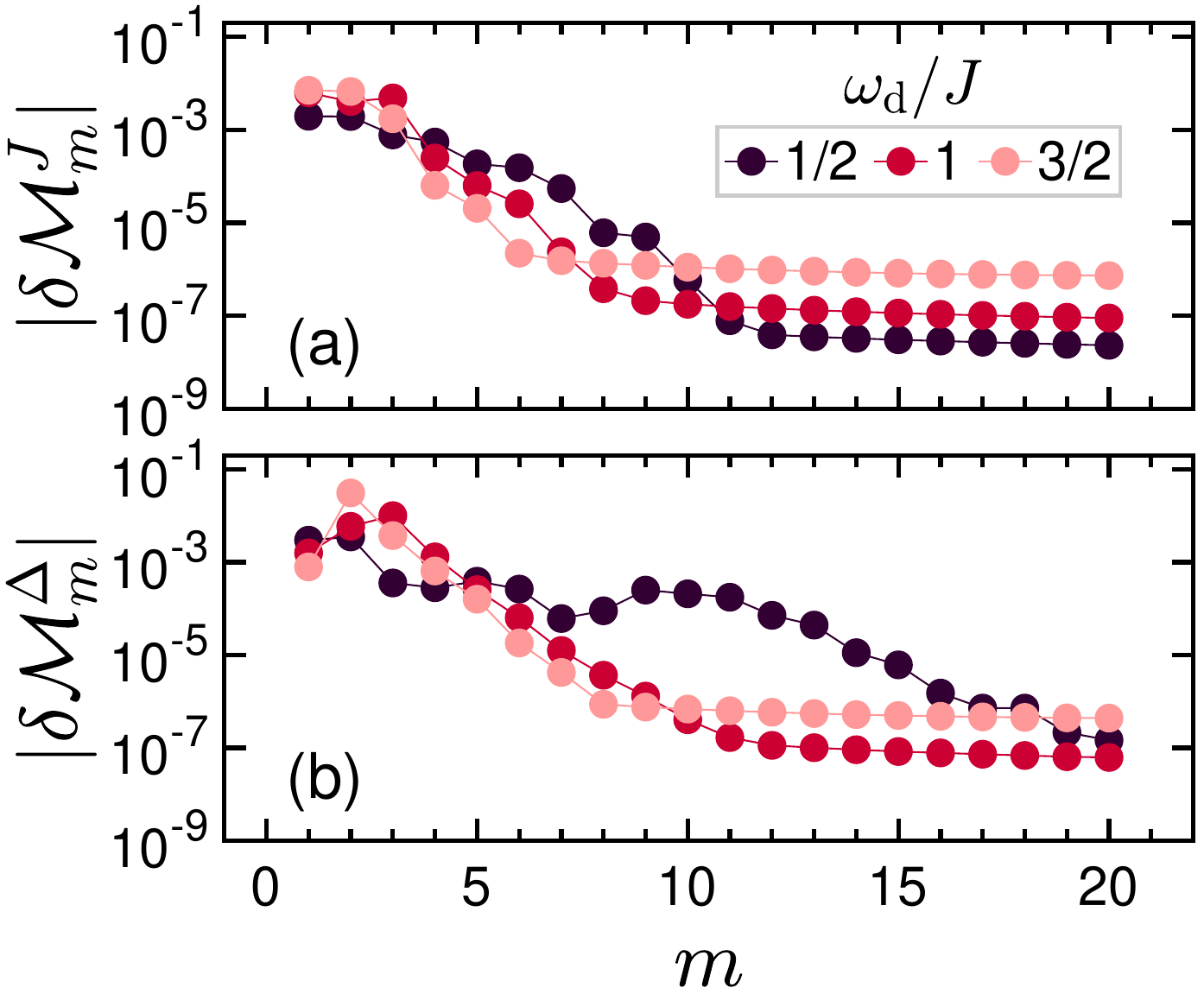}
        \caption{\textbf{Laser drive frequency effect on the HHG spectrum in nonlinear magnetization.} HHG spectrum for various drive frequencies $\omega_{\rm d} = \omega_{\rm p}$ for (a) easy-plane ($J$) and (b) the easy-axis ($\Delta$) interactions in the NESS. The dispersion of the spin band ranges approximately from 1.325$J$ to 2.4$J$ for $\Delta = 1.2$ in our model. Moving closer to the spin band results in a reduction in the number of harmonics for both types of interactions due to resonance effects. \iffalse However, the easy-plane nature of spins generally leads to the generation of a larger number of harmonics.\fi The simulation parameters are $g_{\rm l} = g_{\rm q} = J/4$, $\gamma_{\rm p} = J/40$, $\gamma_{\rm s} = J/100$, and $E_0 = J/50$.} 
		 	\label{f4}
		 \end{figure}

We observe that both odd and even multiples of the driving frequency appear in the harmonic spectrum under linear SPC, while only even harmonics emerge under quadratic SPC. Quadratic SPC results in even harmonics due to the inversion-symmetrical nature of the interaction, leading to an even inversion-symmetry in the system's response. This differs from linear coupling, which can yield both even and odd harmonics depending on specific interaction characteristics. Additionally, small kinks in the HHG spectrum are footprints of the magnon band edges. For the chosen anisotropy parameter $\Delta = 1.2$ throughout this paper, the magnon band disperses approximately from 1.325$J$ to 2.4$J$. For instance, when considering $m = 5$ for $\omega_{\rm d} = J/2$ in Figs.~\ref{f3}(a) and~\ref{f3}(b) for the linear SPC, $\omega = 2.5 J$ corresponds to a frequency near the upper edge of the one magnon band.  \begin{figure}[b]
		 	\centering
		 	\includegraphics[width=1\linewidth]{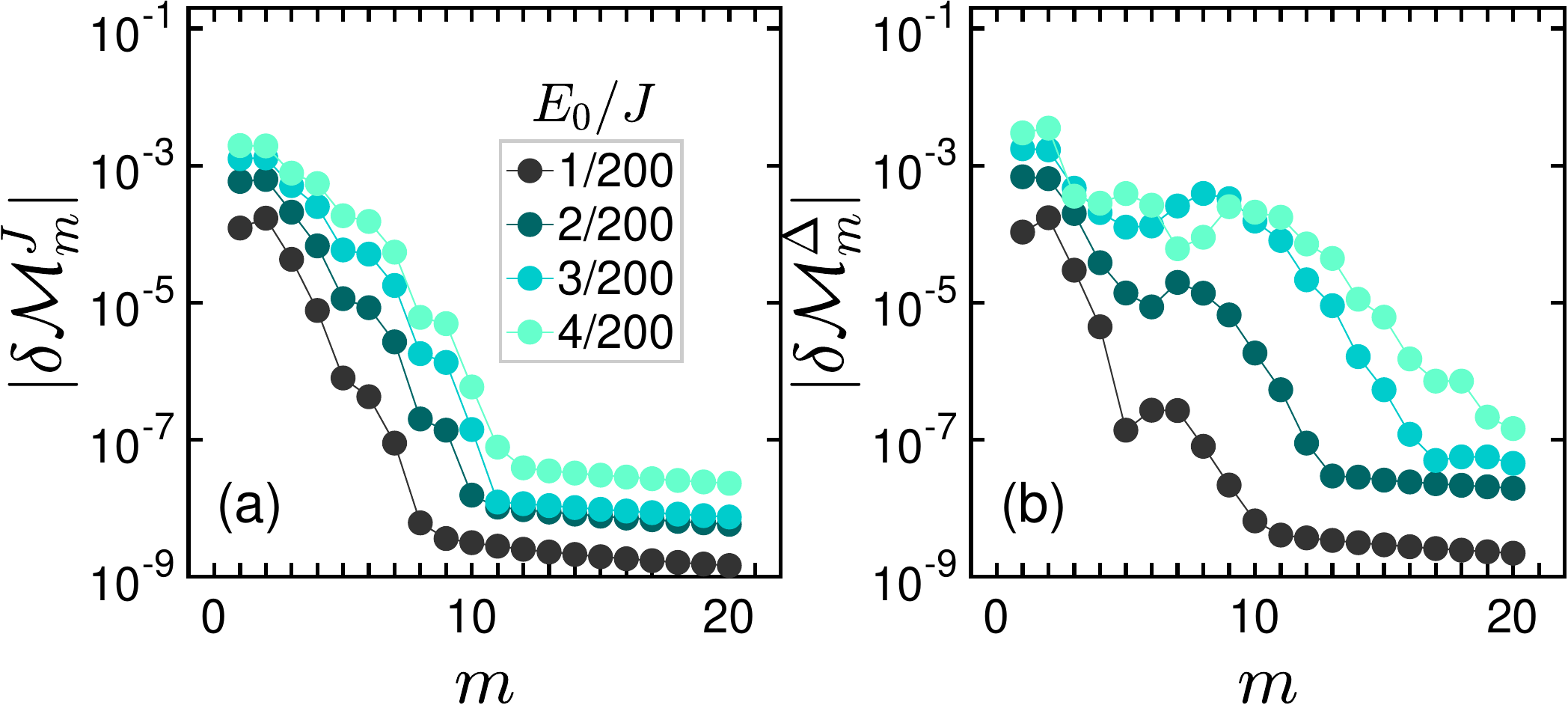}
        \caption{\textbf{Laser drive amplitude effect on the HHG spectrum in nonlinear magnetization.} HHG spectrum for various drive amplitudes $E_0$ under (a) easy-plane ($J$) and (b) the easy-axis ($\Delta$) interactions in the NESS. As the laser amplitude decreases, there is a corresponding reduction in the number of harmonics, with a more pronounced impact observed for the easy-plane interaction. \iffalse Solid and dashed lines in panel (c) show the power-law scaling of harmonics with $E_0$.\fi The fixed parameters are $\omega_{\rm d} = \omega_{\rm p} = J/2$, $g_{\rm l} = g_{\rm q} = J/4$, $\gamma_{\rm p} = J/40$, and $\gamma_{\rm s} = J/100$.} 
		 	\label{f5}
		 \end{figure}

To better understand the dependence on material parameters, Fig.~\ref{f4} depicts the evolution of harmonics under the influence of various drive frequencies, while also modifying the phonon frequency to keep the drive and phonon in resonance. For both the easy-plane and the easy-axis interactions, we again find a cascade of HHG. At drive frequencies $\omega_{\rm d} = \omega_{\rm p} = J/2$ far from the spin frequencies~(dispersing approximately from $1.325 J$ to $2.4 J$), the laser-driven phonon and spins do not resonate, and more harmonics can be generated. As the drive frequency moves near or located in the resonance zone~(for $\omega_{\rm d} = \omega_{\rm p} = J$ and $3J/2$), the number of harmonics decreases as most of the input laser power can be absorbed by the resonantly spin-coupled phonon and spins do not respond nonlinearly. 

It is also worth exploring the possibility of generating more harmonics with laser amplitude $E_0$. In Fig.~\ref{f5}, by applying a stronger laser field, we find that the number of harmonics is effectively increased. This is rooted in Fermi's golden rule, where adjusting the strength of the driving field can effectively enhance or reduce the system's response within the framework of perturbation theory. We note that the laser amplitudes obey the Lindemann criterion~\cite{Lindemann1910} to take care of the lattice melting issue.\begin{figure}[t]
		 	\centering
		 	\includegraphics[width=0.95\linewidth]{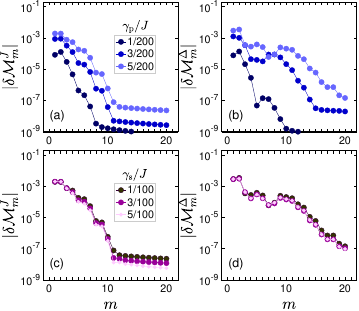}
        \caption{\textbf{Phonon and spin damping effects on the HHG in nonlinear magnetization.} HHG spectrum for various dampings $\gamma_{\rm p}$ for \{(a),(c)\} easy-plane ($J$) and \{(b),(d)\} the easy-axis ($\Delta$) interactions in the NESS. In contrast to spin damping with negligible effects on harmonics, increasing the phonon damping increases the number of harmonics as it brings the phonons back into resonance, allowing stronger phonon excitation. \iffalse, with a stronger effect observed for the easy-plane interaction.\fi We fix the parameters at $\omega_{\rm d} = \omega_{\rm p} = J/2$, $g_{\rm l} = g_{\rm q} = J/4$, $\gamma_{\rm s} = J/100$, and $E_0 = J/50$.} 
		 	\label{f6}
		 \end{figure}

Finally, we focus on the behavior of the coupled system with the damping of both spins and phonons. It is crucial to emphasize that allowable perturbations cannot reach arbitrarily high intensities. In the case of strong phonon damping depicted in Figs.~\ref{f6}(a) and~\ref{f6}(b), our simulations indicate an increase in the number of harmonics in both interactions. This observation can be explained by the direct driving of phonons by the laser. With increased damping in the phonon sector, the effectiveness of SPCs weakens, meaning that strong phonon damping brings the phonons back into resonance, allowing stronger phonon excitation, which leads to an increase in the number of harmonics in the HHG spectrum. In contrast, since the laser indirectly drives spins, strong damping conditions in the spin system result in a partial decoupling of spins from the lattice. As a result, the number of harmonics remains unchanged with spin damping, as depicted in Figs.~\ref{f6}(c) and~\ref{f6}(d). This phenomenon varies across magnetic materials, depending on the source of nonlinear effects. For instance, in a spin-1/2 chain with dimerization, the effect of both phonon and spin dampings, though expected to be finite, are negligible~\cite{allafi2024spin}.

Although our work focuses on the theoretical realm of HHG in ultrafast nonlinear magnetization of a gapped antiferromagnetic spin chain, experimental techniques like time-resolved magneto-optical Kerr effect~\cite{Kimel_2022,PhysRevResearch.4.L022062,gray2024timeresolved} and pump-probe spectroscopy~\cite{PhysRevB.107.094501,https://doi.org/10.1002/qute.202100052} can be employed to measure the magnetization dynamics of such systems. A THz pump laser is typically required to deliver sufficient energy to alter the magnetic properties of the sample effectively. Additionally, the wavelength and intensity of the pump laser must align with the absorption features of the sample material. Through the observation of alterations in optical properties such as reflectivity, transmittance, or Faraday rotation induced by magnetization dynamics, these methods offer valuable insights into the ultrafast phenomena in antiferromagnetic materials. Theoretical models like the Drude-Lorentz model, the Fresnel equations, or specific magneto-optical models are often employed to analyze these changes. 

The other experimental issue regards our choice of laser drive, which involved an instantaneous quench of a monochromatic drive field at time $t=0$. Actual experiments will always involve ramps of this intensity on timescales that are limited by the THz field generation. Initial simulations of such ramps indicate a propensity to instability similar to what is found in other parts of parameter space, but in a manner that depends on the ramp timescale. A more detailed investigation of this ramp dependence will be an interesting topic for future work.

\section{Conclusions}\label{s4}

 Magnonics and spintronics research fields are currently highly sought after since using magnetic excitations for information transport and processing is particularly appealing. In this regard, the nonlinear dynamics of magnetization in antiferromagnetic spin systems hold significant importance. The potential to trigger high-harmonic generation in the non-equilibrium steady state of magnetization presents an exciting opportunity for rapid magnetic state switching in spintronic applications. However, there is still a crucial need for exploring various fundamental mechanisms driving nonlinear spin dynamics in antiferromagnetic materials. Among different ways to generate higher harmonics in antiferromagnets, previous research has suggested that THz electric fields do not induce higher-order harmonic oscillations in the magnetization of antiferromagnets at low temperatures. However, we leverage a THz steady electric field to \textit{indirectly} drive spins via infrared-active phonons and then incorporate both linear and quadratic spin-phonon couplings (SPCs) to elucidate the resulting nonlinear dynamics and generate higher harmonics. In doing so, we employ a combination of spin-wave theory, mean-field theory, and Lindblad formalism. 
 
 Multi-harmonics emerge in the ultrafast magnetization dynamics of a gapped antiferromagnetic chain. Interestingly, the presence of linear~(quadratic) SPC facilitates the generation of both odd and even~(only even) harmonics due to the symmetry nature of couplings. Furthermore, our findings underscore the distinct mechanisms of harmonic generation associated with the easy-plane and the easy-axis interactions.
The outcomes of these investigations hold promise for the development of groundbreaking spin-based devices with improved capabilities.

			\section*{Acknowledgments}
			M.Y. greatly thanks Peter M. Oppeneer for helpful discussions. This work was performed with support from the National Science Foundation (NSF) through award numbers MPS-2228725 and DMR-1945529 and the Welch Foundation through award number AT-2036-20200401 (MK and MY). Part of this work was performed at the Aspen Center for Physics, which is supported by NSF grant No. PHY-1607611, and at the Kavli Institute for Theoretical Physics, which is supported by NSF grant No. NSF PHY-1748958. This project was funded by The University of Texas at Dallas Office of Research and Innovation through the SPIRe program.
 
\appendix
\section{Off-resonant laser-phonon coupling}\label{apa}
In this Appendix, we demonstrate the effects of setting the laser frequency $\omega_{\rm d}$ off-resonance with respect to the phonon frequency $\omega_{\rm p}$. Specifically, in Fig.~\ref{f8}, we explore both redshift and blueshift scenarios, wherein the laser frequency is either lower or higher than the phonon frequency, respectively. In these off-resonance cases, we observe a reduction in the number of harmonics generated and a decrease in their amplitudes compared to the resonance scenario where the laser frequency matches the phonon frequency ($\omega_{\rm d} = \omega_{\rm p}$). This finding emphasizes the significance of resonance states in maximizing the efficiency of energy transfer between the laser and phonon, leading to enhanced nonlinear responses in the system. Although in the off-resonance regime, a stronger electric field might yield additional harmonics, we stick to the resonance case throughout the paper.\begin{figure}[b]
		 	\centering
		 	\includegraphics[width=0.95\linewidth]{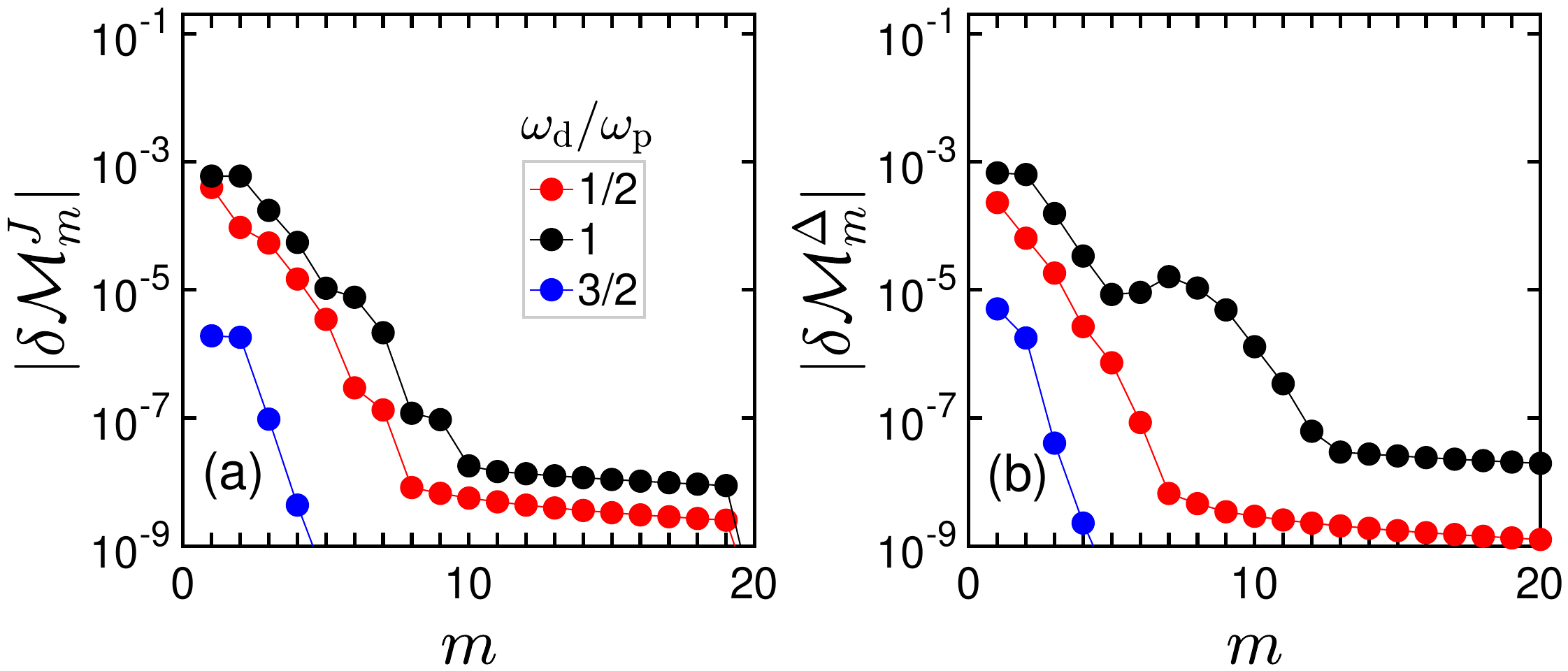}
        \caption{\textbf{Off-resonant laser-phonon coupling effect on the HHG in nonlinear magnetization.} HHG spectrum for off-resonant laser-phonon coupling under (a) easy-plane ($J$) and (b) the easy-axis ($\Delta$) interactions in the NESS. With both redshift and blueshift scenarios, there is a corresponding reduction in the number and amplitude of harmonics. The fixed parameters are $\omega_{\rm d} = \omega_{\rm p} = J/2$, $g_{\rm l} = g_{\rm q} = J/4$, $\gamma_{\rm p} = J/40$, $\gamma_{\rm s} = J/100$, and $E_0 = J/100$.} 
		 	\label{f8}
		 \end{figure} 
}
	\bibliography{bib}

\begin{thebibliography}{79}%
\makeatletter
\providecommand \@ifxundefined [1]{%
 \@ifx{#1\undefined}
}%
\providecommand \@ifnum [1]{%
 \ifnum #1\expandafter \@firstoftwo
 \else \expandafter \@secondoftwo
 \fi
}%
\providecommand \@ifx [1]{%
 \ifx #1\expandafter \@firstoftwo
 \else \expandafter \@secondoftwo
 \fi
}%
\providecommand \natexlab [1]{#1}%
\providecommand \enquote  [1]{``#1''}%
\providecommand \bibnamefont  [1]{#1}%
\providecommand \bibfnamefont [1]{#1}%
\providecommand \citenamefont [1]{#1}%
\providecommand \href@noop [0]{\@secondoftwo}%
\providecommand \href [0]{\begingroup \@sanitize@url \@href}%
\providecommand \@href[1]{\@@startlink{#1}\@@href}%
\providecommand \@@href[1]{\endgroup#1\@@endlink}%
\providecommand \@sanitize@url [0]{\catcode `\\12\catcode `\$12\catcode
  `\&12\catcode `\#12\catcode `\^12\catcode `\_12\catcode `\%12\relax}%
\providecommand \@@startlink[1]{}%
\providecommand \@@endlink[0]{}%
\providecommand \url  [0]{\begingroup\@sanitize@url \@url }%
\providecommand \@url [1]{\endgroup\@href {#1}{\urlprefix }}%
\providecommand \urlprefix  [0]{URL }%
\providecommand \Eprint [0]{\href }%
\providecommand \doibase [0]{http://dx.doi.org/}%
\providecommand \selectlanguage [0]{\@gobble}%
\providecommand \bibinfo  [0]{\@secondoftwo}%
\providecommand \bibfield  [0]{\@secondoftwo}%
\providecommand \translation [1]{[#1]}%
\providecommand \BibitemOpen [0]{}%
\providecommand \bibitemStop [0]{}%
\providecommand \bibitemNoStop [0]{.\EOS\space}%
\providecommand \EOS [0]{\spacefactor3000\relax}%
\providecommand \BibitemShut  [1]{\csname bibitem#1\endcsname}%
\let\auto@bib@innerbib\@empty
\bibitem [{\citenamefont {\ifmmode \check{Z}\else
  \v{Z}\fi{}uti\ifmmode~\acute{c}\else \'{c}\fi{}}\ \emph
  {et~al.}(2004)\citenamefont {\ifmmode \check{Z}\else
  \v{Z}\fi{}uti\ifmmode~\acute{c}\else \'{c}\fi{}}, \citenamefont {Fabian},\
  and\ \citenamefont {Das~Sarma}}]{RevModPhys.76.323}%
  \BibitemOpen
  \bibfield  {author} {\bibinfo {author} {\bibfnamefont {I.}~\bibnamefont
  {\ifmmode \check{Z}\else \v{Z}\fi{}uti\ifmmode~\acute{c}\else \'{c}\fi{}}},
  \bibinfo {author} {\bibfnamefont {J.}~\bibnamefont {Fabian}}, \ and\ \bibinfo
  {author} {\bibfnamefont {S.}~\bibnamefont {Das~Sarma}},\ }\href {\doibase
  10.1103/RevModPhys.76.323} {\bibfield  {journal} {\bibinfo  {journal} {Rev.
  Mod. Phys.}\ }\textbf {\bibinfo {volume} {76}},\ \bibinfo {pages} {323}
  (\bibinfo {year} {2004})}\BibitemShut {NoStop}%
\bibitem [{\citenamefont {Hirohata}\ \emph {et~al.}(2020)\citenamefont
  {Hirohata}, \citenamefont {Yamada}, \citenamefont {Nakatani}, \citenamefont
  {Prejbeanu}, \citenamefont {Diény}, \citenamefont {Pirro},\ and\
  \citenamefont {Hillebrands}}]{HIROHATA2020166711}%
  \BibitemOpen
  \bibfield  {author} {\bibinfo {author} {\bibfnamefont {A.}~\bibnamefont
  {Hirohata}}, \bibinfo {author} {\bibfnamefont {K.}~\bibnamefont {Yamada}},
  \bibinfo {author} {\bibfnamefont {Y.}~\bibnamefont {Nakatani}}, \bibinfo
  {author} {\bibfnamefont {I.-L.}\ \bibnamefont {Prejbeanu}}, \bibinfo {author}
  {\bibfnamefont {B.}~\bibnamefont {Diény}}, \bibinfo {author} {\bibfnamefont
  {P.}~\bibnamefont {Pirro}}, \ and\ \bibinfo {author} {\bibfnamefont
  {B.}~\bibnamefont {Hillebrands}},\ }\href {\doibase
  https://doi.org/10.1016/j.jmmm.2020.166711} {\bibfield  {journal} {\bibinfo
  {journal} {Journal of Magnetism and Magnetic Materials}\ }\textbf {\bibinfo
  {volume} {509}},\ \bibinfo {pages} {166711} (\bibinfo {year}
  {2020})}\BibitemShut {NoStop}%
\bibitem [{\citenamefont {Yuan}\ \emph {et~al.}(2022)\citenamefont {Yuan},
  \citenamefont {Cao}, \citenamefont {Kamra}, \citenamefont {Duine},\ and\
  \citenamefont {Yan}}]{YUAN20221}%
  \BibitemOpen
  \bibfield  {author} {\bibinfo {author} {\bibfnamefont {H.}~\bibnamefont
  {Yuan}}, \bibinfo {author} {\bibfnamefont {Y.}~\bibnamefont {Cao}}, \bibinfo
  {author} {\bibfnamefont {A.}~\bibnamefont {Kamra}}, \bibinfo {author}
  {\bibfnamefont {R.~A.}\ \bibnamefont {Duine}}, \ and\ \bibinfo {author}
  {\bibfnamefont {P.}~\bibnamefont {Yan}},\ }\href {\doibase
  https://doi.org/10.1016/j.physrep.2022.03.002} {\bibfield  {journal}
  {\bibinfo  {journal} {Physics Reports}\ }\textbf {\bibinfo {volume} {965}},\
  \bibinfo {pages} {1} (\bibinfo {year} {2022})}\BibitemShut {NoStop}%
\bibitem [{\citenamefont {Hirohata}(2022)}]{magnetochemistry8040037}%
  \BibitemOpen
  \bibfield  {author} {\bibinfo {author} {\bibfnamefont {A.}~\bibnamefont
  {Hirohata}},\ }\href {\doibase 10.3390/magnetochemistry8040037} {\bibfield
  {journal} {\bibinfo  {journal} {Magnetochemistry}\ }\textbf {\bibinfo
  {volume} {8}} (\bibinfo {year} {2022}),\
  10.3390/magnetochemistry8040037}\BibitemShut {NoStop}%
\bibitem [{\citenamefont {Chumak}\ \emph {et~al.}(2015)\citenamefont {Chumak},
  \citenamefont {Vasyuchka}, \citenamefont {Serga},\ and\ \citenamefont
  {Hillebrands}}]{Chumak2015}%
  \BibitemOpen
  \bibfield  {author} {\bibinfo {author} {\bibfnamefont {A.~V.}\ \bibnamefont
  {Chumak}}, \bibinfo {author} {\bibfnamefont {V.~.~I.}\ \bibnamefont
  {Vasyuchka}}, \bibinfo {author} {\bibfnamefont {A.~.~A.}\ \bibnamefont
  {Serga}}, \ and\ \bibinfo {author} {\bibfnamefont {B.}~\bibnamefont
  {Hillebrands}}}\href {\doibase 10.1038/nphys3347} {\bibfield  {journal}
  {\bibinfo  {journal} {Nature Physics}\ }\textbf {\bibinfo {volume} {11}},\
  \bibinfo {pages} {453} (\bibinfo {year} {2015})}\BibitemShut {NoStop}%
\bibitem [{\citenamefont {Jungwirth}\ \emph {et~al.}(2016)\citenamefont
  {Jungwirth}, \citenamefont {Marti}, \citenamefont {Wadley},\ and\
  \citenamefont {Wunderlich}}]{Jungwirth2016}%
  \BibitemOpen
  \bibfield  {author} {\bibinfo {author} {\bibfnamefont {T.}~\bibnamefont
  {Jungwirth}}, \bibinfo {author} {\bibfnamefont {X.}~\bibnamefont {Marti}},
  \bibinfo {author} {\bibfnamefont {P.}~\bibnamefont {Wadley}}, \ and\ \bibinfo
  {author} {\bibfnamefont {J.}~\bibnamefont {Wunderlich}},\ }\href {\doibase
  10.1038/nnano.2016.18} {\bibfield  {journal} {\bibinfo  {journal} {Nature
  Nanotechnology}\ }\textbf {\bibinfo {volume} {11}},\ \bibinfo {pages} {231}
  (\bibinfo {year} {2016})}\BibitemShut {NoStop}%
\bibitem [{\citenamefont {Baltz}\ \emph {et~al.}(2018)\citenamefont {Baltz},
  \citenamefont {Manchon}, \citenamefont {Tsoi}, \citenamefont {Moriyama},
  \citenamefont {Ono},\ and\ \citenamefont
  {Tserkovnyak}}]{RevModPhys.90.015005}%
  \BibitemOpen
  \bibfield  {author} {\bibinfo {author} {\bibfnamefont {V.}~\bibnamefont
  {Baltz}}, \bibinfo {author} {\bibfnamefont {A.}~\bibnamefont {Manchon}},
  \bibinfo {author} {\bibfnamefont {M.}~\bibnamefont {Tsoi}}, \bibinfo {author}
  {\bibfnamefont {T.}~\bibnamefont {Moriyama}}, \bibinfo {author}
  {\bibfnamefont {T.}~\bibnamefont {Ono}}, \ and\ \bibinfo {author}
  {\bibfnamefont {Y.}~\bibnamefont {Tserkovnyak}},\ }\href {\doibase
  10.1103/RevModPhys.90.015005} {\bibfield  {journal} {\bibinfo  {journal}
  {Rev. Mod. Phys.}\ }\textbf {\bibinfo {volume} {90}},\ \bibinfo {pages}
  {015005} (\bibinfo {year} {2018})}\BibitemShut {NoStop}%
\bibitem [{\citenamefont {Uhrig}(2024)}]{uhrig2024landaulifshitz}%
  \BibitemOpen
  \bibfield  {author} {\bibinfo {author} {\bibfnamefont {G.~S.}\ \bibnamefont
  {Uhrig}},\ }\href@noop {} {} (\bibinfo {year} {2024}),\ \Eprint
  {http://arxiv.org/abs/2406.10613} {arXiv:2406.10613 [cond-mat.str-el]}
  \BibitemShut {NoStop}%
\bibitem [{\citenamefont {Bolsmann}\ \emph {et~al.}(2023)\citenamefont
  {Bolsmann}, \citenamefont {Khudoyberdiev},\ and\ \citenamefont
  {Uhrig}}]{PRXQuantum.4.030332}%
  \BibitemOpen
  \bibfield  {author} {\bibinfo {author} {\bibfnamefont {K.}~\bibnamefont
  {Bolsmann}}, \bibinfo {author} {\bibfnamefont {A.}~\bibnamefont
  {Khudoyberdiev}}, \ and\ \bibinfo {author} {\bibfnamefont {G.~S.}\
  \bibnamefont {Uhrig}},\ }\href {\doibase 10.1103/PRXQuantum.4.030332}
  {\bibfield  {journal} {\bibinfo  {journal} {PRX Quantum}\ }\textbf {\bibinfo
  {volume} {4}},\ \bibinfo {pages} {030332} (\bibinfo {year}
  {2023})}\BibitemShut {NoStop}%
\bibitem [{\citenamefont {Khudoyberdiev}\ and\ \citenamefont
  {Uhrig}(2024)}]{PhysRevB.109.174419}%
  \BibitemOpen
  \bibfield  {author} {\bibinfo {author} {\bibfnamefont {A.}~\bibnamefont
  {Khudoyberdiev}}\ and\ \bibinfo {author} {\bibfnamefont {G.~S.}\ \bibnamefont
  {Uhrig}},\ }\href {\doibase 10.1103/PhysRevB.109.174419} {\bibfield
  {journal} {\bibinfo  {journal} {Phys. Rev. B}\ }\textbf {\bibinfo {volume}
  {109}},\ \bibinfo {pages} {174419} (\bibinfo {year} {2024})}\BibitemShut
  {NoStop}%
\bibitem [{\citenamefont {Bossini}\ \emph {et~al.}(2021)\citenamefont
  {Bossini}, \citenamefont {Dal~Conte}, \citenamefont {Terschanski},
  \citenamefont {Springholz}, \citenamefont {Bonanni}, \citenamefont
  {Deltenre}, \citenamefont {Anders}, \citenamefont {Uhrig}, \citenamefont
  {Cerullo},\ and\ \citenamefont {Cinchetti}}]{PhysRevB.104.224424}%
  \BibitemOpen
  \bibfield  {author} {\bibinfo {author} {\bibfnamefont {D.}~\bibnamefont
  {Bossini}}, \bibinfo {author} {\bibfnamefont {S.}~\bibnamefont {Dal~Conte}},
  \bibinfo {author} {\bibfnamefont {M.}~\bibnamefont {Terschanski}}, \bibinfo
  {author} {\bibfnamefont {G.}~\bibnamefont {Springholz}}, \bibinfo {author}
  {\bibfnamefont {A.}~\bibnamefont {Bonanni}}, \bibinfo {author} {\bibfnamefont
  {K.}~\bibnamefont {Deltenre}}, \bibinfo {author} {\bibfnamefont
  {F.}~\bibnamefont {Anders}}, \bibinfo {author} {\bibfnamefont {G.~S.}\
  \bibnamefont {Uhrig}}, \bibinfo {author} {\bibfnamefont {G.}~\bibnamefont
  {Cerullo}}, \ and\ \bibinfo {author} {\bibfnamefont {M.}~\bibnamefont
  {Cinchetti}},\ }\href {\doibase 10.1103/PhysRevB.104.224424} {\bibfield
  {journal} {\bibinfo  {journal} {Phys. Rev. B}\ }\textbf {\bibinfo {volume}
  {104}},\ \bibinfo {pages} {224424} (\bibinfo {year} {2021})}\BibitemShut
  {NoStop}%
\bibitem [{\citenamefont {Bossini}\ \emph {et~al.}(2020)\citenamefont
  {Bossini}, \citenamefont {Terschanski}, \citenamefont {Mertens},
  \citenamefont {Springholz}, \citenamefont {Bonanni}, \citenamefont {Uhrig},\
  and\ \citenamefont {Cinchetti}}]{Bossini_2020}%
  \BibitemOpen
  \bibfield  {author} {\bibinfo {author} {\bibfnamefont {D.}~\bibnamefont
  {Bossini}}, \bibinfo {author} {\bibfnamefont {M.}~\bibnamefont
  {Terschanski}}, \bibinfo {author} {\bibfnamefont {F.}~\bibnamefont
  {Mertens}}, \bibinfo {author} {\bibfnamefont {G.}~\bibnamefont {Springholz}},
  \bibinfo {author} {\bibfnamefont {A.}~\bibnamefont {Bonanni}}, \bibinfo
  {author} {\bibfnamefont {G.~S.}\ \bibnamefont {Uhrig}}, \ and\ \bibinfo
  {author} {\bibfnamefont {M.}~\bibnamefont {Cinchetti}},\ }\href {\doibase
  10.1088/1367-2630/aba0e7} {\bibfield  {journal} {\bibinfo  {journal} {New
  Journal of Physics}\ }\textbf {\bibinfo {volume} {22}},\ \bibinfo {pages}
  {083029} (\bibinfo {year} {2020})}\BibitemShut {NoStop}%
\bibitem [{\citenamefont {Deltenre}\ \emph {et~al.}(2021)\citenamefont
  {Deltenre}, \citenamefont {Bossini}, \citenamefont {Anders},\ and\
  \citenamefont {Uhrig}}]{PhysRevB.104.184419}%
  \BibitemOpen
  \bibfield  {author} {\bibinfo {author} {\bibfnamefont {K.}~\bibnamefont
  {Deltenre}}, \bibinfo {author} {\bibfnamefont {D.}~\bibnamefont {Bossini}},
  \bibinfo {author} {\bibfnamefont {F.~B.}\ \bibnamefont {Anders}}, \ and\
  \bibinfo {author} {\bibfnamefont {G.~S.}\ \bibnamefont {Uhrig}},\ }\href
  {\doibase 10.1103/PhysRevB.104.184419} {\bibfield  {journal} {\bibinfo
  {journal} {Phys. Rev. B}\ }\textbf {\bibinfo {volume} {104}},\ \bibinfo
  {pages} {184419} (\bibinfo {year} {2021})}\BibitemShut {NoStop}%
\bibitem [{\citenamefont {Saitoh}\ \emph {et~al.}(2006)\citenamefont {Saitoh},
  \citenamefont {Ueda}, \citenamefont {Miyajima},\ and\ \citenamefont
  {Tatara}}]{10.1063/1.2199473}%
  \BibitemOpen
  \bibfield  {author} {\bibinfo {author} {\bibfnamefont {E.}~\bibnamefont
  {Saitoh}}, \bibinfo {author} {\bibfnamefont {M.}~\bibnamefont {Ueda}},
  \bibinfo {author} {\bibfnamefont {H.}~\bibnamefont {Miyajima}}, \ and\
  \bibinfo {author} {\bibfnamefont {G.}~\bibnamefont {Tatara}},\ }\href
  {\doibase 10.1063/1.2199473} {\bibfield  {journal} {\bibinfo  {journal}
  {Applied Physics Letters}\ }\textbf {\bibinfo {volume} {88}},\ \bibinfo
  {pages} {182509} (\bibinfo {year} {2006})}\BibitemShut {NoStop}%
\bibitem [{\citenamefont {Wadley}\ \emph {et~al.}(2016)\citenamefont {Wadley},
  \citenamefont {Howells}, \citenamefont {Železný}, \citenamefont {Andrews},
  \citenamefont {Hills}, \citenamefont {Campion}, \citenamefont {Novák},
  \citenamefont {Olejník}, \citenamefont {Maccherozzi}, \citenamefont {Dhesi}
  \emph {et~al.}}]{doi:10.1126/science.aab1031}%
  \BibitemOpen
  \bibfield  {author} {\bibinfo {author} {\bibfnamefont {P.}~\bibnamefont
  {Wadley}}, \bibinfo {author} {\bibfnamefont {B.}~\bibnamefont {Howells}},
  \bibinfo {author} {\bibfnamefont {J.}~\bibnamefont {Železný}}, \bibinfo
  {author} {\bibfnamefont {C.}~\bibnamefont {Andrews}}, \bibinfo {author}
  {\bibfnamefont {V.}~\bibnamefont {Hills}}, \bibinfo {author} {\bibfnamefont
  {R.~P.}\ \bibnamefont {Campion}}, \bibinfo {author} {\bibfnamefont
  {V.}~\bibnamefont {Novák}}, \bibinfo {author} {\bibfnamefont
  {K.}~\bibnamefont {Olejník}}, \bibinfo {author} {\bibfnamefont
  {F.}~\bibnamefont {Maccherozzi}}, \bibinfo {author} {\bibfnamefont {S.~S.}\
  \bibnamefont {Dhesi}},  \emph {et~al.},\ }\href {\doibase
  10.1126/science.aab1031} {\bibfield  {journal} {\bibinfo  {journal}
  {Science}\ }\textbf {\bibinfo {volume} {351}},\ \bibinfo {pages} {587}
  (\bibinfo {year} {2016})}\BibitemShut {NoStop}%
\bibitem [{\citenamefont {N{\v{e}}mec}\ \emph {et~al.}(2018)\citenamefont
  {N{\v{e}}mec}, \citenamefont {Fiebig}, \citenamefont {Kampfrath},\ and\
  \citenamefont {Kimel}}]{Nemec2018}%
  \BibitemOpen
  \bibfield  {author} {\bibinfo {author} {\bibfnamefont {P.}~\bibnamefont
  {N{\v{e}}mec}}, \bibinfo {author} {\bibfnamefont {M.}~\bibnamefont {Fiebig}},
  \bibinfo {author} {\bibfnamefont {T.}~\bibnamefont {Kampfrath}}, \ and\
  \bibinfo {author} {\bibfnamefont {A.~V.}\ \bibnamefont {Kimel}},\ }\href
  {\doibase 10.1038/s41567-018-0051-x} {\bibfield  {journal} {\bibinfo
  {journal} {Nature Physics}\ }\textbf {\bibinfo {volume} {14}},\ \bibinfo
  {pages} {229} (\bibinfo {year} {2018})}\BibitemShut {NoStop}%
\bibitem [{\citenamefont {Kampfrath}\ \emph {et~al.}(2011)\citenamefont
  {Kampfrath}, \citenamefont {Sell}, \citenamefont {Klatt}, \citenamefont
  {Pashkin}, \citenamefont {M{\"a}hrlein}, \citenamefont {Dekorsy},
  \citenamefont {Wolf}, \citenamefont {Fiebig}, \citenamefont {Leitenstorfer},\
  and\ \citenamefont {Huber}}]{Kampfrath2011}%
  \BibitemOpen
  \bibfield  {author} {\bibinfo {author} {\bibfnamefont {T.}~\bibnamefont
  {Kampfrath}}, \bibinfo {author} {\bibfnamefont {A.}~\bibnamefont {Sell}},
  \bibinfo {author} {\bibfnamefont {G.}~\bibnamefont {Klatt}}, \bibinfo
  {author} {\bibfnamefont {A.}~\bibnamefont {Pashkin}}, \bibinfo {author}
  {\bibfnamefont {S.}~\bibnamefont {M{\"a}hrlein}}, \bibinfo {author}
  {\bibfnamefont {T.}~\bibnamefont {Dekorsy}}, \bibinfo {author} {\bibfnamefont
  {M.}~\bibnamefont {Wolf}}, \bibinfo {author} {\bibfnamefont {M.}~\bibnamefont
  {Fiebig}}, \bibinfo {author} {\bibfnamefont {A.}~\bibnamefont
  {Leitenstorfer}}, \ and\ \bibinfo {author} {\bibfnamefont {R.}~\bibnamefont
  {Huber}},\ }\href {\doibase 10.1038/nphoton.2010.259} {\bibfield  {journal}
  {\bibinfo  {journal} {Nature Photonics}\ }\textbf {\bibinfo {volume} {5}},\
  \bibinfo {pages} {31} (\bibinfo {year} {2011})}\BibitemShut {NoStop}%
\bibitem [{\citenamefont {Olejník}\ \emph {et~al.}(2018)\citenamefont
  {Olejník}, \citenamefont {Seifert}, \citenamefont {Kašpar}, \citenamefont
  {Novák}, \citenamefont {Wadley}, \citenamefont {Campion}, \citenamefont
  {Baumgartner}, \citenamefont {Gambardella}, \citenamefont {Němec},
  \citenamefont {Wunderlich} \emph {et~al.}}]{doi:10.1126/sciadv.aar3566}%
  \BibitemOpen
  \bibfield  {author} {\bibinfo {author} {\bibfnamefont {K.}~\bibnamefont
  {Olejník}}, \bibinfo {author} {\bibfnamefont {T.}~\bibnamefont {Seifert}},
  \bibinfo {author} {\bibfnamefont {Z.}~\bibnamefont {Kašpar}}, \bibinfo
  {author} {\bibfnamefont {V.}~\bibnamefont {Novák}}, \bibinfo {author}
  {\bibfnamefont {P.}~\bibnamefont {Wadley}}, \bibinfo {author} {\bibfnamefont
  {R.~P.}\ \bibnamefont {Campion}}, \bibinfo {author} {\bibfnamefont
  {M.}~\bibnamefont {Baumgartner}}, \bibinfo {author} {\bibfnamefont
  {P.}~\bibnamefont {Gambardella}}, \bibinfo {author} {\bibfnamefont
  {P.}~\bibnamefont {Němec}}, \bibinfo {author} {\bibfnamefont
  {J.}~\bibnamefont {Wunderlich}},  \emph {et~al.},\ }\href {\doibase
  10.1126/sciadv.aar3566} {\bibfield  {journal} {\bibinfo  {journal} {Science
  Advances}\ }\textbf {\bibinfo {volume} {4}},\ \bibinfo {pages} {eaar3566}
  (\bibinfo {year} {2018})}\BibitemShut {NoStop}%
\bibitem [{\citenamefont {Li}\ \emph {et~al.}(2020)\citenamefont {Li},
  \citenamefont {Wilson}, \citenamefont {Cheng}, \citenamefont {Lohmann},
  \citenamefont {Kavand}, \citenamefont {Yuan}, \citenamefont {Aldosary},
  \citenamefont {Agladze}, \citenamefont {Wei}, \citenamefont {Sherwin} \emph
  {et~al.}}]{Li2020}%
  \BibitemOpen
  \bibfield  {author} {\bibinfo {author} {\bibfnamefont {J.}~\bibnamefont
  {Li}}, \bibinfo {author} {\bibfnamefont {C.~B.}\ \bibnamefont {Wilson}},
  \bibinfo {author} {\bibfnamefont {R.}~\bibnamefont {Cheng}}, \bibinfo
  {author} {\bibfnamefont {M.}~\bibnamefont {Lohmann}}, \bibinfo {author}
  {\bibfnamefont {M.}~\bibnamefont {Kavand}}, \bibinfo {author} {\bibfnamefont
  {W.}~\bibnamefont {Yuan}}, \bibinfo {author} {\bibfnamefont {M.}~\bibnamefont
  {Aldosary}}, \bibinfo {author} {\bibfnamefont {N.}~\bibnamefont {Agladze}},
  \bibinfo {author} {\bibfnamefont {P.}~\bibnamefont {Wei}}, \bibinfo {author}
  {\bibfnamefont {M.~S.}\ \bibnamefont {Sherwin}},  \emph {et~al.},\ }\href
  {\doibase 10.1038/s41586-020-1950-4} {\bibfield  {journal} {\bibinfo
  {journal} {Nature}\ }\textbf {\bibinfo {volume} {578}},\ \bibinfo {pages}
  {70} (\bibinfo {year} {2020})}\BibitemShut {NoStop}%
\bibitem [{\citenamefont {Vaidya}\ \emph {et~al.}(2020)\citenamefont {Vaidya},
  \citenamefont {Morley}, \citenamefont {van Tol}, \citenamefont {Liu},
  \citenamefont {Cheng}, \citenamefont {Brataas}, \citenamefont {Lederman},\
  and\ \citenamefont {del Barco}}]{doi:10.1126/science.aaz4247}%
  \BibitemOpen
  \bibfield  {author} {\bibinfo {author} {\bibfnamefont {P.}~\bibnamefont
  {Vaidya}}, \bibinfo {author} {\bibfnamefont {S.~A.}\ \bibnamefont {Morley}},
  \bibinfo {author} {\bibfnamefont {J.}~\bibnamefont {van Tol}}, \bibinfo
  {author} {\bibfnamefont {Y.}~\bibnamefont {Liu}}, \bibinfo {author}
  {\bibfnamefont {R.}~\bibnamefont {Cheng}}, \bibinfo {author} {\bibfnamefont
  {A.}~\bibnamefont {Brataas}}, \bibinfo {author} {\bibfnamefont
  {D.}~\bibnamefont {Lederman}}, \ and\ \bibinfo {author} {\bibfnamefont
  {E.}~\bibnamefont {del Barco}},\ }\href {\doibase 10.1126/science.aaz4247}
  {\bibfield  {journal} {\bibinfo  {journal} {Science}\ }\textbf {\bibinfo
  {volume} {368}},\ \bibinfo {pages} {160} (\bibinfo {year}
  {2020})}\BibitemShut {NoStop}%
\bibitem [{\citenamefont {Li}\ \emph {et~al.}(2022)\citenamefont {Li},
  \citenamefont {Yang}, \citenamefont {Mondal}, \citenamefont {Tzschaschel},\
  and\ \citenamefont {Pal}}]{10.1063/5.0075999}%
  \BibitemOpen
  \bibfield  {author} {\bibinfo {author} {\bibfnamefont {J.}~\bibnamefont
  {Li}}, \bibinfo {author} {\bibfnamefont {C.-J.}\ \bibnamefont {Yang}},
  \bibinfo {author} {\bibfnamefont {R.}~\bibnamefont {Mondal}}, \bibinfo
  {author} {\bibfnamefont {C.}~\bibnamefont {Tzschaschel}}, \ and\ \bibinfo
  {author} {\bibfnamefont {S.}~\bibnamefont {Pal}},\ }\href {\doibase
  10.1063/5.0075999} {\bibfield  {journal} {\bibinfo  {journal} {Applied
  Physics Letters}\ }\textbf {\bibinfo {volume} {120}},\ \bibinfo {pages}
  {050501} (\bibinfo {year} {2022})}\BibitemShut {NoStop}%
\bibitem [{\citenamefont {Mrudul}\ \emph {et~al.}(2020)\citenamefont {Mrudul},
  \citenamefont {Tancogne-Dejean}, \citenamefont {Rubio},\ and\ \citenamefont
  {Dixit}}]{Mrudul2020}%
  \BibitemOpen
  \bibfield  {author} {\bibinfo {author} {\bibfnamefont {M.~S.}\ \bibnamefont
  {Mrudul}}, \bibinfo {author} {\bibfnamefont {N.}~\bibnamefont
  {Tancogne-Dejean}}, \bibinfo {author} {\bibfnamefont {A.}~\bibnamefont
  {Rubio}}, \ and\ \bibinfo {author} {\bibfnamefont {G.}~\bibnamefont
  {Dixit}},\ }\href {\doibase 10.1038/s41524-020-0275-z} {\bibfield  {journal}
  {\bibinfo  {journal} {npj Computational Materials}\ }\textbf {\bibinfo
  {volume} {6}},\ \bibinfo {pages} {10} (\bibinfo {year} {2020})}\BibitemShut
  {NoStop}%
\bibitem [{\citenamefont {Mrudul}\ and\ \citenamefont
  {Oppeneer}(2024)}]{PhysRevB.109.144418}%
  \BibitemOpen
  \bibfield  {author} {\bibinfo {author} {\bibfnamefont {M.~S.}\ \bibnamefont
  {Mrudul}}\ and\ \bibinfo {author} {\bibfnamefont {P.~M.}\ \bibnamefont
  {Oppeneer}},\ }\href {\doibase 10.1103/PhysRevB.109.144418} {\bibfield
  {journal} {\bibinfo  {journal} {Phys. Rev. B}\ }\textbf {\bibinfo {volume}
  {109}},\ \bibinfo {pages} {144418} (\bibinfo {year} {2024})}\BibitemShut
  {NoStop}%
\bibitem [{\citenamefont {Gupta}\ \emph {et~al.}(2023)\citenamefont {Gupta},
  \citenamefont {Cosco}, \citenamefont {Malik}, \citenamefont {Chen},
  \citenamefont {Saha}, \citenamefont {Ghosh}, \citenamefont {Pohlmann},
  \citenamefont {Mardegan}, \citenamefont {Francoual}, \citenamefont
  {Stefanuik}, \citenamefont {S\"oderstr\"om}, \citenamefont {Sanyal},
  \citenamefont {Karis}, \citenamefont {Svedlindh}, \citenamefont {Oppeneer},\
  and\ \citenamefont {Knut}}]{PhysRevB.108.064427}%
  \BibitemOpen
  \bibfield  {author} {\bibinfo {author} {\bibfnamefont {R.}~\bibnamefont
  {Gupta}}, \bibinfo {author} {\bibfnamefont {F.}~\bibnamefont {Cosco}},
  \bibinfo {author} {\bibfnamefont {R.~S.}\ \bibnamefont {Malik}}, \bibinfo
  {author} {\bibfnamefont {X.}~\bibnamefont {Chen}}, \bibinfo {author}
  {\bibfnamefont {S.}~\bibnamefont {Saha}}, \bibinfo {author} {\bibfnamefont
  {A.}~\bibnamefont {Ghosh}}, \bibinfo {author} {\bibfnamefont
  {T.}~\bibnamefont {Pohlmann}}, \bibinfo {author} {\bibfnamefont {J.~R.~L.}\
  \bibnamefont {Mardegan}}, \bibinfo {author} {\bibfnamefont {S.}~\bibnamefont
  {Francoual}}, \bibinfo {author} {\bibfnamefont {R.}~\bibnamefont
  {Stefanuik}}, \bibinfo {author} {\bibfnamefont {J.}~\bibnamefont
  {S\"oderstr\"om}}, \bibinfo {author} {\bibfnamefont {B.}~\bibnamefont
  {Sanyal}}, \bibinfo {author} {\bibfnamefont {O.}~\bibnamefont {Karis}},
  \bibinfo {author} {\bibfnamefont {P.}~\bibnamefont {Svedlindh}}, \bibinfo
  {author} {\bibfnamefont {P.~M.}\ \bibnamefont {Oppeneer}}, \ and\ \bibinfo
  {author} {\bibfnamefont {R.}~\bibnamefont {Knut}},\ }\href {\doibase
  10.1103/PhysRevB.108.064427} {\bibfield  {journal} {\bibinfo  {journal}
  {Phys. Rev. B}\ }\textbf {\bibinfo {volume} {108}},\ \bibinfo {pages}
  {064427} (\bibinfo {year} {2023})}\BibitemShut {NoStop}%
\bibitem [{\citenamefont {Zhang}\ \emph {et~al.}(2023)\citenamefont {Zhang},
  \citenamefont {Sekiguchi}, \citenamefont {Moriyama}, \citenamefont {Furuya},
  \citenamefont {Sato}, \citenamefont {Satoh}, \citenamefont {Mukai},
  \citenamefont {Tanaka}, \citenamefont {Yamamoto} \emph {et~al.}}]{Zhang2023}%
  \BibitemOpen
  \bibfield  {author} {\bibinfo {author} {\bibfnamefont {Z.}~\bibnamefont
  {Zhang}}, \bibinfo {author} {\bibfnamefont {F.}~\bibnamefont {Sekiguchi}},
  \bibinfo {author} {\bibfnamefont {T.}~\bibnamefont {Moriyama}}, \bibinfo
  {author} {\bibfnamefont {S.~C.}\ \bibnamefont {Furuya}}, \bibinfo {author}
  {\bibfnamefont {M.}~\bibnamefont {Sato}}, \bibinfo {author} {\bibfnamefont
  {T.}~\bibnamefont {Satoh}}, \bibinfo {author} {\bibfnamefont
  {Y.}~\bibnamefont {Mukai}}, \bibinfo {author} {\bibfnamefont
  {K.}~\bibnamefont {Tanaka}}, \bibinfo {author} {\bibfnamefont
  {T.}~\bibnamefont {Yamamoto}},  \emph {et~al.},\ }\href {\doibase
  10.1038/s41467-023-37473-1} {\bibfield  {journal} {\bibinfo  {journal}
  {Nature Communications}\ }\textbf {\bibinfo {volume} {14}},\ \bibinfo {pages}
  {1795} (\bibinfo {year} {2023})}\BibitemShut {NoStop}%
\bibitem [{\citenamefont {Schlauderer}\ \emph {et~al.}(2019)\citenamefont
  {Schlauderer}, \citenamefont {Lange}, \citenamefont {Baierl}, \citenamefont
  {Ebnet}, \citenamefont {Schmid}, \citenamefont {Valovcin}, \citenamefont
  {Zvezdin}, \citenamefont {Kimel}, \citenamefont {Mikhaylovskiy},\ and\
  \citenamefont {Huber}}]{Schlauderer2019}%
  \BibitemOpen
  \bibfield  {author} {\bibinfo {author} {\bibfnamefont {S.}~\bibnamefont
  {Schlauderer}}, \bibinfo {author} {\bibfnamefont {C.}~\bibnamefont {Lange}},
  \bibinfo {author} {\bibfnamefont {S.}~\bibnamefont {Baierl}}, \bibinfo
  {author} {\bibfnamefont {T.}~\bibnamefont {Ebnet}}, \bibinfo {author}
  {\bibfnamefont {C.~P.}\ \bibnamefont {Schmid}}, \bibinfo {author}
  {\bibfnamefont {D.~C.}\ \bibnamefont {Valovcin}}, \bibinfo {author}
  {\bibfnamefont {A.~K.}\ \bibnamefont {Zvezdin}}, \bibinfo {author}
  {\bibfnamefont {A.~V.}\ \bibnamefont {Kimel}}, \bibinfo {author}
  {\bibfnamefont {R.~V.}\ \bibnamefont {Mikhaylovskiy}}, \ and\ \bibinfo
  {author} {\bibfnamefont {R.}~\bibnamefont {Huber}},\ }\href {\doibase
  10.1038/s41586-019-1174-7} {\bibfield  {journal} {\bibinfo  {journal}
  {Nature}\ }\textbf {\bibinfo {volume} {569}},\ \bibinfo {pages} {383}
  (\bibinfo {year} {2019})}\BibitemShut {NoStop}%
\bibitem [{\citenamefont {Hart}\ \emph {et~al.}(2024)\citenamefont {Hart},
  \citenamefont {Sutcliffe}, \citenamefont {Refael},\ and\ \citenamefont
  {Paramekanti}}]{hart2024phonondriven}%
  \BibitemOpen
  \bibfield  {author} {\bibinfo {author} {\bibfnamefont {K.}~\bibnamefont
  {Hart}}, \bibinfo {author} {\bibfnamefont {R.}~\bibnamefont {Sutcliffe}},
  \bibinfo {author} {\bibfnamefont {G.}~\bibnamefont {Refael}}, \ and\ \bibinfo
  {author} {\bibfnamefont {A.}~\bibnamefont {Paramekanti}},\ }\href@noop {} {}
  (\bibinfo {year} {2024}),\ \Eprint {http://arxiv.org/abs/2404.17633}
  {arXiv:2404.17633 [cond-mat.str-el]} \BibitemShut {NoStop}%
\bibitem [{\citenamefont {Wang}\ and\ \citenamefont
  {Vishwanath}(2008)}]{PhysRevLett.100.077201}%
  \BibitemOpen
  \bibfield  {author} {\bibinfo {author} {\bibfnamefont {F.}~\bibnamefont
  {Wang}}\ and\ \bibinfo {author} {\bibfnamefont {A.}~\bibnamefont
  {Vishwanath}},\ }\href {\doibase 10.1103/PhysRevLett.100.077201} {\bibfield
  {journal} {\bibinfo  {journal} {Phys. Rev. Lett.}\ }\textbf {\bibinfo
  {volume} {100}},\ \bibinfo {pages} {077201} (\bibinfo {year}
  {2008})}\BibitemShut {NoStop}%
\bibitem [{\citenamefont {Hernandez}\ \emph {et~al.}(2023)\citenamefont
  {Hernandez}, \citenamefont {Baydin}, \citenamefont {Chaudhary}, \citenamefont
  {Tay}, \citenamefont {Katayama}, \citenamefont {Takeda}, \citenamefont
  {Nojiri}, \citenamefont {Okazaki}, \citenamefont {Rappl}, \citenamefont
  {Abramof} \emph {et~al.}}]{Hernandez_2023}%
  \BibitemOpen
  \bibfield  {author} {\bibinfo {author} {\bibfnamefont {F.~G.}\ \bibnamefont
  {Hernandez}}, \bibinfo {author} {\bibfnamefont {A.}~\bibnamefont {Baydin}},
  \bibinfo {author} {\bibfnamefont {S.}~\bibnamefont {Chaudhary}}, \bibinfo
  {author} {\bibfnamefont {F.}~\bibnamefont {Tay}}, \bibinfo {author}
  {\bibfnamefont {I.}~\bibnamefont {Katayama}}, \bibinfo {author}
  {\bibfnamefont {J.}~\bibnamefont {Takeda}}, \bibinfo {author} {\bibfnamefont
  {H.}~\bibnamefont {Nojiri}}, \bibinfo {author} {\bibfnamefont {A.~K.}\
  \bibnamefont {Okazaki}}, \bibinfo {author} {\bibfnamefont {P.~H.}\
  \bibnamefont {Rappl}}, \bibinfo {author} {\bibfnamefont {E.}~\bibnamefont
  {Abramof}},  \emph {et~al.},\ }\href {\doibase 10.1126/sciadv.adj4074}
  {\bibfield  {journal} {\bibinfo  {journal} {Science Advances}\ }\textbf
  {\bibinfo {volume} {9}} (\bibinfo {year} {2023}),\
  10.1126/sciadv.adj4074}\BibitemShut {NoStop}%
\bibitem [{\citenamefont {Nieves}\ \emph {et~al.}(2016)\citenamefont {Nieves},
  \citenamefont {Serantes},\ and\ \citenamefont
  {Chubykalo-Fesenko}}]{PhysRevB.94.014409}%
  \BibitemOpen
  \bibfield  {author} {\bibinfo {author} {\bibfnamefont {P.}~\bibnamefont
  {Nieves}}, \bibinfo {author} {\bibfnamefont {D.}~\bibnamefont {Serantes}}, \
  and\ \bibinfo {author} {\bibfnamefont {O.}~\bibnamefont
  {Chubykalo-Fesenko}},\ }\href {\doibase 10.1103/PhysRevB.94.014409}
  {\bibfield  {journal} {\bibinfo  {journal} {Phys. Rev. B}\ }\textbf {\bibinfo
  {volume} {94}},\ \bibinfo {pages} {014409} (\bibinfo {year}
  {2016})}\BibitemShut {NoStop}%
\bibitem [{\citenamefont {Juraschek}\ \emph {et~al.}(2022)\citenamefont
  {Juraschek}, \citenamefont {Neuman},\ and\ \citenamefont
  {Narang}}]{PhysRevResearch.4.013129}%
  \BibitemOpen
  \bibfield  {author} {\bibinfo {author} {\bibfnamefont {D.~M.}\ \bibnamefont
  {Juraschek}}, \bibinfo {author} {\bibfnamefont {T.~c.~v.}\ \bibnamefont
  {Neuman}}, \ and\ \bibinfo {author} {\bibfnamefont {P.}~\bibnamefont
  {Narang}},\ }\href {\doibase 10.1103/PhysRevResearch.4.013129} {\bibfield
  {journal} {\bibinfo  {journal} {Phys. Rev. Res.}\ }\textbf {\bibinfo {volume}
  {4}},\ \bibinfo {pages} {013129} (\bibinfo {year} {2022})}\BibitemShut
  {NoStop}%
\bibitem [{\citenamefont {Curtis}\ \emph {et~al.}(2023)\citenamefont {Curtis},
  \citenamefont {Disa}, \citenamefont {Fechner}, \citenamefont {Cavalleri},\
  and\ \citenamefont {Narang}}]{PhysRevResearch.5.013204}%
  \BibitemOpen
  \bibfield  {author} {\bibinfo {author} {\bibfnamefont {J.~B.}\ \bibnamefont
  {Curtis}}, \bibinfo {author} {\bibfnamefont {A.}~\bibnamefont {Disa}},
  \bibinfo {author} {\bibfnamefont {M.}~\bibnamefont {Fechner}}, \bibinfo
  {author} {\bibfnamefont {A.}~\bibnamefont {Cavalleri}}, \ and\ \bibinfo
  {author} {\bibfnamefont {P.}~\bibnamefont {Narang}},\ }\href {\doibase
  10.1103/PhysRevResearch.5.013204} {\bibfield  {journal} {\bibinfo  {journal}
  {Phys. Rev. Res.}\ }\textbf {\bibinfo {volume} {5}},\ \bibinfo {pages}
  {013204} (\bibinfo {year} {2023})}\BibitemShut {NoStop}%
\bibitem [{\citenamefont {Yarmohammadi}\ \emph
  {et~al.}(2024{\natexlab{a}})\citenamefont {Yarmohammadi}, \citenamefont
  {Bukov}, \citenamefont {Oganesyan},\ and\ \citenamefont
  {Kolodrubetz}}]{PhysRevB.109.224417}%
  \BibitemOpen
  \bibfield  {author} {\bibinfo {author} {\bibfnamefont {M.}~\bibnamefont
  {Yarmohammadi}}, \bibinfo {author} {\bibfnamefont {M.}~\bibnamefont {Bukov}},
  \bibinfo {author} {\bibfnamefont {V.}~\bibnamefont {Oganesyan}}, \ and\
  \bibinfo {author} {\bibfnamefont {M.~H.}\ \bibnamefont {Kolodrubetz}},\
  }\href {\doibase 10.1103/PhysRevB.109.224417} {\bibfield  {journal} {\bibinfo
   {journal} {Phys. Rev. B}\ }\textbf {\bibinfo {volume} {109}},\ \bibinfo
  {pages} {224417} (\bibinfo {year} {2024}{\natexlab{a}})}\BibitemShut
  {NoStop}%
\bibitem [{\citenamefont {Allafi}\ \emph {et~al.}(2024)\citenamefont {Allafi},
  \citenamefont {Kolodrubetz}, \citenamefont {Bukov}, \citenamefont
  {Oganesyan},\ and\ \citenamefont {Yarmohammadi}}]{allafi2024spin}%
  \BibitemOpen
  \bibfield  {author} {\bibinfo {author} {\bibfnamefont {N.~M.}\ \bibnamefont
  {Allafi}}, \bibinfo {author} {\bibfnamefont {M.~H.}\ \bibnamefont
  {Kolodrubetz}}, \bibinfo {author} {\bibfnamefont {M.}~\bibnamefont {Bukov}},
  \bibinfo {author} {\bibfnamefont {V.}~\bibnamefont {Oganesyan}}, \ and\
  \bibinfo {author} {\bibfnamefont {M.}~\bibnamefont {Yarmohammadi}},\
  }\href@noop {} {} (\bibinfo {year} {2024}),\ \Eprint
  {http://arxiv.org/abs/2404.05830} {arXiv:2404.05830 [cond-mat.str-el]}
  \BibitemShut {NoStop}%
\bibitem [{\citenamefont {Strungaru}\ \emph {et~al.}(2023)\citenamefont
  {Strungaru}, \citenamefont {Ellis}, \citenamefont {Ruta}, \citenamefont
  {Evans}, \citenamefont {Chantrell},\ and\ \citenamefont
  {Chubykalo-Fesenko}}]{strungaru2023route}%
  \BibitemOpen
  \bibfield  {author} {\bibinfo {author} {\bibfnamefont {M.}~\bibnamefont
  {Strungaru}}, \bibinfo {author} {\bibfnamefont {M.~O.~A.}\ \bibnamefont
  {Ellis}}, \bibinfo {author} {\bibfnamefont {S.}~\bibnamefont {Ruta}},
  \bibinfo {author} {\bibfnamefont {R.~F.~L.}\ \bibnamefont {Evans}}, \bibinfo
  {author} {\bibfnamefont {R.~W.}\ \bibnamefont {Chantrell}}, \ and\ \bibinfo
  {author} {\bibfnamefont {O.}~\bibnamefont {Chubykalo-Fesenko}},\ }\href@noop
  {} {} (\bibinfo {year} {2023}),\ \Eprint {http://arxiv.org/abs/2209.04312}
  {arXiv:2209.04312 [cond-mat.mtrl-sci]} \BibitemShut {NoStop}%
\bibitem [{\citenamefont {Zhou}\ \emph {et~al.}(2024)\citenamefont {Zhou},
  \citenamefont {Li}, \citenamefont {Frauenheim},\ and\ \citenamefont
  {He}}]{zhou2024coherent}%
  \BibitemOpen
  \bibfield  {author} {\bibinfo {author} {\bibfnamefont {Z.}~\bibnamefont
  {Zhou}}, \bibinfo {author} {\bibfnamefont {M.}~\bibnamefont {Li}}, \bibinfo
  {author} {\bibfnamefont {T.}~\bibnamefont {Frauenheim}}, \ and\ \bibinfo
  {author} {\bibfnamefont {J.}~\bibnamefont {He}},\ }\href@noop {} {} (\bibinfo
  {year} {2024}),\ \Eprint {http://arxiv.org/abs/2403.15204} {arXiv:2403.15204
  [physics.comp-ph]} \BibitemShut {NoStop}%
\bibitem [{\citenamefont {Winta}\ \emph {et~al.}(2018)\citenamefont {Winta},
  \citenamefont {Gewinner}, \citenamefont {Sch\"ollkopf}, \citenamefont
  {Wolf},\ and\ \citenamefont {Paarmann}}]{PhysRevB.97.094108}%
  \BibitemOpen
  \bibfield  {author} {\bibinfo {author} {\bibfnamefont {C.~J.}\ \bibnamefont
  {Winta}}, \bibinfo {author} {\bibfnamefont {S.}~\bibnamefont {Gewinner}},
  \bibinfo {author} {\bibfnamefont {W.}~\bibnamefont {Sch\"ollkopf}}, \bibinfo
  {author} {\bibfnamefont {M.}~\bibnamefont {Wolf}}, \ and\ \bibinfo {author}
  {\bibfnamefont {A.}~\bibnamefont {Paarmann}},\ }\href {\doibase
  10.1103/PhysRevB.97.094108} {\bibfield  {journal} {\bibinfo  {journal} {Phys.
  Rev. B}\ }\textbf {\bibinfo {volume} {97}},\ \bibinfo {pages} {094108}
  (\bibinfo {year} {2018})}\BibitemShut {NoStop}%
\bibitem [{\citenamefont {Ginsberg}\ \emph {et~al.}(2023)\citenamefont
  {Ginsberg}, \citenamefont {Jadidi}, \citenamefont {Zhang}, \citenamefont
  {Chen}, \citenamefont {Tancogne-Dejean}, \citenamefont {Chae}, \citenamefont
  {Patwardhan}, \citenamefont {Xian}, \citenamefont {Watanabe}, \citenamefont
  {Taniguchi}, \citenamefont {Hone}, \citenamefont {Rubio},\ and\ \citenamefont
  {Gaeta}}]{Ginsberg2023}%
  \BibitemOpen
  \bibfield  {author} {\bibinfo {author} {\bibfnamefont {J.~S.}\ \bibnamefont
  {Ginsberg}}, \bibinfo {author} {\bibfnamefont {M.~M.}\ \bibnamefont
  {Jadidi}}, \bibinfo {author} {\bibfnamefont {J.}~\bibnamefont {Zhang}},
  \bibinfo {author} {\bibfnamefont {C.~Y.}\ \bibnamefont {Chen}}, \bibinfo
  {author} {\bibfnamefont {N.}~\bibnamefont {Tancogne-Dejean}}, \bibinfo
  {author} {\bibfnamefont {S.~H.}\ \bibnamefont {Chae}}, \bibinfo {author}
  {\bibfnamefont {G.~N.}\ \bibnamefont {Patwardhan}}, \bibinfo {author}
  {\bibfnamefont {L.}~\bibnamefont {Xian}}, \bibinfo {author} {\bibfnamefont
  {K.}~\bibnamefont {Watanabe}}, \bibinfo {author} {\bibfnamefont
  {T.}~\bibnamefont {Taniguchi}}, \bibinfo {author} {\bibfnamefont
  {J.}~\bibnamefont {Hone}}, \bibinfo {author} {\bibfnamefont {A.}~\bibnamefont
  {Rubio}}, \ and\ \bibinfo {author} {\bibfnamefont {A.~L.}\ \bibnamefont
  {Gaeta}},\ }\href {\doibase 10.1038/s41467-023-43501-x} {\bibfield  {journal}
  {\bibinfo  {journal} {Nature Communications}\ }\textbf {\bibinfo {volume}
  {14}},\ \bibinfo {pages} {7685} (\bibinfo {year} {2023})}\BibitemShut
  {NoStop}%
\bibitem [{\citenamefont {Rana}\ \emph {et~al.}(2022)\citenamefont {Rana},
  \citenamefont {Mrudul}, \citenamefont {Kartashov}, \citenamefont {Ivanov},\
  and\ \citenamefont {Dixit}}]{PhysRevB.106.064303}%
  \BibitemOpen
  \bibfield  {author} {\bibinfo {author} {\bibfnamefont {N.}~\bibnamefont
  {Rana}}, \bibinfo {author} {\bibfnamefont {M.~S.}\ \bibnamefont {Mrudul}},
  \bibinfo {author} {\bibfnamefont {D.}~\bibnamefont {Kartashov}}, \bibinfo
  {author} {\bibfnamefont {M.}~\bibnamefont {Ivanov}}, \ and\ \bibinfo {author}
  {\bibfnamefont {G.}~\bibnamefont {Dixit}},\ }\href {\doibase
  10.1103/PhysRevB.106.064303} {\bibfield  {journal} {\bibinfo  {journal}
  {Phys. Rev. B}\ }\textbf {\bibinfo {volume} {106}},\ \bibinfo {pages}
  {064303} (\bibinfo {year} {2022})}\BibitemShut {NoStop}%
\bibitem [{\citenamefont {Zheng}\ \emph {et~al.}(2024)\citenamefont {Zheng},
  \citenamefont {Khalsa},\ and\ \citenamefont
  {Moses}}]{zheng2024phononmediated}%
  \BibitemOpen
  \bibfield  {author} {\bibinfo {author} {\bibfnamefont {J.}~\bibnamefont
  {Zheng}}, \bibinfo {author} {\bibfnamefont {G.}~\bibnamefont {Khalsa}}, \
  and\ \bibinfo {author} {\bibfnamefont {J.}~\bibnamefont {Moses}},\
  }\href@noop {} {} (\bibinfo {year} {2024}),\ \Eprint
  {http://arxiv.org/abs/2309.00167} {arXiv:2309.00167 [physics.optics]}
  \BibitemShut {NoStop}%
\bibitem [{\citenamefont {Rana}\ and\ \citenamefont
  {Dixit}(2022)}]{PhysRevA.106.053116}%
  \BibitemOpen
  \bibfield  {author} {\bibinfo {author} {\bibfnamefont {N.}~\bibnamefont
  {Rana}}\ and\ \bibinfo {author} {\bibfnamefont {G.}~\bibnamefont {Dixit}},\
  }\href {\doibase 10.1103/PhysRevA.106.053116} {\bibfield  {journal} {\bibinfo
   {journal} {Phys. Rev. A}\ }\textbf {\bibinfo {volume} {106}},\ \bibinfo
  {pages} {053116} (\bibinfo {year} {2022})}\BibitemShut {NoStop}%
\bibitem [{\citenamefont {Zhang}\ \emph {et~al.}(2018)\citenamefont {Zhang},
  \citenamefont {Si}, \citenamefont {Murakami}, \citenamefont {Bai},\ and\
  \citenamefont {George}}]{zhang2018generating}%
  \BibitemOpen
  \bibfield  {author} {\bibinfo {author} {\bibfnamefont {G.~P.}\ \bibnamefont
  {Zhang}}, \bibinfo {author} {\bibfnamefont {M.~S.}\ \bibnamefont {Si}},
  \bibinfo {author} {\bibfnamefont {M.}~\bibnamefont {Murakami}}, \bibinfo
  {author} {\bibfnamefont {Y.~H.}\ \bibnamefont {Bai}}, \ and\ \bibinfo
  {author} {\bibfnamefont {T.~F.}\ \bibnamefont {George}},\ }\href {\doibase
  10.1038/s41467-018-05535-4} {\bibfield  {journal} {\bibinfo  {journal}
  {Nature Communications}\ }\textbf {\bibinfo {volume} {9}},\ \bibinfo {pages}
  {3031} (\bibinfo {year} {2018})}\BibitemShut {NoStop}%
\bibitem [{\citenamefont {Lysne}\ \emph {et~al.}(2020)\citenamefont {Lysne},
  \citenamefont {Murakami}, \citenamefont {Sch\"uler},\ and\ \citenamefont
  {Werner}}]{PhysRevB.102.081121}%
  \BibitemOpen
  \bibfield  {author} {\bibinfo {author} {\bibfnamefont {M.}~\bibnamefont
  {Lysne}}, \bibinfo {author} {\bibfnamefont {Y.}~\bibnamefont {Murakami}},
  \bibinfo {author} {\bibfnamefont {M.}~\bibnamefont {Sch\"uler}}, \ and\
  \bibinfo {author} {\bibfnamefont {P.}~\bibnamefont {Werner}},\ }\href
  {\doibase 10.1103/PhysRevB.102.081121} {\bibfield  {journal} {\bibinfo
  {journal} {Phys. Rev. B}\ }\textbf {\bibinfo {volume} {102}},\ \bibinfo
  {pages} {081121} (\bibinfo {year} {2020})}\BibitemShut {NoStop}%
\bibitem [{\citenamefont {Takayoshi}\ \emph {et~al.}(2019)\citenamefont
  {Takayoshi}, \citenamefont {Murakami},\ and\ \citenamefont
  {Werner}}]{PhysRevB.99.184303}%
  \BibitemOpen
  \bibfield  {author} {\bibinfo {author} {\bibfnamefont {S.}~\bibnamefont
  {Takayoshi}}, \bibinfo {author} {\bibfnamefont {Y.}~\bibnamefont {Murakami}},
  \ and\ \bibinfo {author} {\bibfnamefont {P.}~\bibnamefont {Werner}},\ }\href
  {\doibase 10.1103/PhysRevB.99.184303} {\bibfield  {journal} {\bibinfo
  {journal} {Phys. Rev. B}\ }\textbf {\bibinfo {volume} {99}},\ \bibinfo
  {pages} {184303} (\bibinfo {year} {2019})}\BibitemShut {NoStop}%
\bibitem [{\citenamefont {Kanega}\ \emph {et~al.}(2021)\citenamefont {Kanega},
  \citenamefont {Ikeda},\ and\ \citenamefont {Sato}}]{kanega2021linear}%
  \BibitemOpen
  \bibfield  {author} {\bibinfo {author} {\bibfnamefont {M.}~\bibnamefont
  {Kanega}}, \bibinfo {author} {\bibfnamefont {T.~N.}\ \bibnamefont {Ikeda}}, \
  and\ \bibinfo {author} {\bibfnamefont {M.}~\bibnamefont {Sato}},\ }\href
  {\doibase 10.1103/PhysRevResearch.3.L032024} {\bibfield  {journal} {\bibinfo
  {journal} {Phys. Rev. Res.}\ }\textbf {\bibinfo {volume} {3}},\ \bibinfo
  {pages} {L032024} (\bibinfo {year} {2021})}\BibitemShut {NoStop}%
\bibitem [{\citenamefont {Zhao}\ \emph {et~al.}(2021)\citenamefont {Zhao},
  \citenamefont {Wang}, \citenamefont {Hong}, \citenamefont {Wang},
  \citenamefont {Han}, \citenamefont {Wang}, \citenamefont {Liu}, \citenamefont
  {Long}, \citenamefont {Wang},\ and\ \citenamefont {Lu}}]{ZHAO2021449}%
  \BibitemOpen
  \bibfield  {author} {\bibinfo {author} {\bibfnamefont {W.}~\bibnamefont
  {Zhao}}, \bibinfo {author} {\bibfnamefont {K.}~\bibnamefont {Wang}}, \bibinfo
  {author} {\bibfnamefont {X.}~\bibnamefont {Hong}}, \bibinfo {author}
  {\bibfnamefont {B.}~\bibnamefont {Wang}}, \bibinfo {author} {\bibfnamefont
  {X.}~\bibnamefont {Han}}, \bibinfo {author} {\bibfnamefont {K.}~\bibnamefont
  {Wang}}, \bibinfo {author} {\bibfnamefont {W.}~\bibnamefont {Liu}}, \bibinfo
  {author} {\bibfnamefont {H.}~\bibnamefont {Long}}, \bibinfo {author}
  {\bibfnamefont {B.}~\bibnamefont {Wang}}, \ and\ \bibinfo {author}
  {\bibfnamefont {P.}~\bibnamefont {Lu}},\ }\href {\doibase
  https://doi.org/10.1016/j.scib.2020.08.043} {\bibfield  {journal} {\bibinfo
  {journal} {Science Bulletin}\ }\textbf {\bibinfo {volume} {66}},\ \bibinfo
  {pages} {449} (\bibinfo {year} {2021})}\BibitemShut {NoStop}%
\bibitem [{\citenamefont {Mukai}\ \emph {et~al.}(2016)\citenamefont {Mukai},
  \citenamefont {Hirori}, \citenamefont {Yamamoto}, \citenamefont {Kageyama},\
  and\ \citenamefont {Tanaka}}]{Mukai_2016}%
  \BibitemOpen
  \bibfield  {author} {\bibinfo {author} {\bibfnamefont {Y.}~\bibnamefont
  {Mukai}}, \bibinfo {author} {\bibfnamefont {H.}~\bibnamefont {Hirori}},
  \bibinfo {author} {\bibfnamefont {T.}~\bibnamefont {Yamamoto}}, \bibinfo
  {author} {\bibfnamefont {H.}~\bibnamefont {Kageyama}}, \ and\ \bibinfo
  {author} {\bibfnamefont {K.}~\bibnamefont {Tanaka}},\ }\href {\doibase
  10.1088/1367-2630/18/1/013045} {\bibfield  {journal} {\bibinfo  {journal}
  {New Journal of Physics}\ }\textbf {\bibinfo {volume} {18}},\ \bibinfo
  {pages} {013045} (\bibinfo {year} {2016})}\BibitemShut {NoStop}%
\bibitem [{\citenamefont {Stremoukhov}\ \emph {et~al.}(2024)\citenamefont
  {Stremoukhov}, \citenamefont {Safin}, \citenamefont {Schippers},
  \citenamefont {Lavrijsen}, \citenamefont {Bal}, \citenamefont {Zeitler},
  \citenamefont {Sadovnikov}, \citenamefont {Kozlova}, \citenamefont {Ilkhchy},
  \citenamefont {Nikitov} \emph {et~al.}}]{STREMOUKHOV2024107377}%
  \BibitemOpen
  \bibfield  {author} {\bibinfo {author} {\bibfnamefont {P.}~\bibnamefont
  {Stremoukhov}}, \bibinfo {author} {\bibfnamefont {A.}~\bibnamefont {Safin}},
  \bibinfo {author} {\bibfnamefont {C.~F.}\ \bibnamefont {Schippers}}, \bibinfo
  {author} {\bibfnamefont {R.}~\bibnamefont {Lavrijsen}}, \bibinfo {author}
  {\bibfnamefont {M.}~\bibnamefont {Bal}}, \bibinfo {author} {\bibfnamefont
  {U.}~\bibnamefont {Zeitler}}, \bibinfo {author} {\bibfnamefont
  {A.}~\bibnamefont {Sadovnikov}}, \bibinfo {author} {\bibfnamefont
  {E.}~\bibnamefont {Kozlova}}, \bibinfo {author} {\bibfnamefont {K.~S.}\
  \bibnamefont {Ilkhchy}}, \bibinfo {author} {\bibfnamefont {S.}~\bibnamefont
  {Nikitov}},  \emph {et~al.},\ }\href {\doibase
  https://doi.org/10.1016/j.rinp.2024.107377} {\bibfield  {journal} {\bibinfo
  {journal} {Results in Physics}\ }\textbf {\bibinfo {volume} {57}},\ \bibinfo
  {pages} {107377} (\bibinfo {year} {2024})}\BibitemShut {NoStop}%
\bibitem [{\citenamefont {Mukai}\ \emph {et~al.}(2015)\citenamefont {Mukai},
  \citenamefont {Hirori}, \citenamefont {Yamamoto}, \citenamefont {Kageyama},\
  and\ \citenamefont {Tanaka}}]{Mukai_15}%
  \BibitemOpen
  \bibfield  {author} {\bibinfo {author} {\bibfnamefont {Y.}~\bibnamefont
  {Mukai}}, \bibinfo {author} {\bibfnamefont {H.}~\bibnamefont {Hirori}},
  \bibinfo {author} {\bibfnamefont {T.}~\bibnamefont {Yamamoto}}, \bibinfo
  {author} {\bibfnamefont {H.}~\bibnamefont {Kageyama}}, \ and\ \bibinfo
  {author} {\bibfnamefont {K.}~\bibnamefont {Tanaka}},\ }in\ \href {\doibase
  10.1364/CLEO_SI.2015.STu2H.1} {\emph {\bibinfo {booktitle} {CLEO: 2015}}}\
  (\bibinfo  {publisher} {Optica Publishing Group},\ \bibinfo {year} {2015})\
  p.\ \bibinfo {pages} {STu2H.1}\BibitemShut {NoStop}%
\bibitem [{\citenamefont {Hayami}\ and\ \citenamefont
  {Yatsushiro}(2022)}]{PhysRevB.106.014420}%
  \BibitemOpen
  \bibfield  {author} {\bibinfo {author} {\bibfnamefont {S.}~\bibnamefont
  {Hayami}}\ and\ \bibinfo {author} {\bibfnamefont {M.}~\bibnamefont
  {Yatsushiro}},\ }\href {\doibase 10.1103/PhysRevB.106.014420} {\bibfield
  {journal} {\bibinfo  {journal} {Phys. Rev. B}\ }\textbf {\bibinfo {volume}
  {106}},\ \bibinfo {pages} {014420} (\bibinfo {year} {2022})}\BibitemShut
  {NoStop}%
\bibitem [{\citenamefont {Shih}\ \emph {et~al.}(2013)\citenamefont {Shih},
  \citenamefont {Cheng},\ and\ \citenamefont {Wu}}]{Shih2013}%
  \BibitemOpen
  \bibfield  {author} {\bibinfo {author} {\bibfnamefont {P.-H.}\ \bibnamefont
  {Shih}}, \bibinfo {author} {\bibfnamefont {C.-L.}\ \bibnamefont {Cheng}}, \
  and\ \bibinfo {author} {\bibfnamefont {S.~Y.}\ \bibnamefont {Wu}},\ }\href
  {\doibase 10.1186/1556-276X-8-398} {\bibfield  {journal} {\bibinfo  {journal}
  {Nanoscale Research Letters}\ }\textbf {\bibinfo {volume} {8}},\ \bibinfo
  {pages} {398} (\bibinfo {year} {2013})}\BibitemShut {NoStop}%
\bibitem [{\citenamefont {Hlubek}\ \emph {et~al.}(2012)\citenamefont {Hlubek},
  \citenamefont {Zotos}, \citenamefont {Singh}, \citenamefont {Saint-Martin},
  \citenamefont {Revcolevschi}, \citenamefont {Büchner},\ and\ \citenamefont
  {Hess}}]{Hlubek_2012}%
  \BibitemOpen
  \bibfield  {author} {\bibinfo {author} {\bibfnamefont {N.}~\bibnamefont
  {Hlubek}}, \bibinfo {author} {\bibfnamefont {X.}~\bibnamefont {Zotos}},
  \bibinfo {author} {\bibfnamefont {S.}~\bibnamefont {Singh}}, \bibinfo
  {author} {\bibfnamefont {R.}~\bibnamefont {Saint-Martin}}, \bibinfo {author}
  {\bibfnamefont {A.}~\bibnamefont {Revcolevschi}}, \bibinfo {author}
  {\bibfnamefont {B.}~\bibnamefont {Büchner}}, \ and\ \bibinfo {author}
  {\bibfnamefont {C.}~\bibnamefont {Hess}},\ }\href {\doibase
  10.1088/1742-5468/2012/03/p03006} {\bibfield  {journal} {\bibinfo  {journal}
  {Journal of Statistical Mechanics: Theory and Experiment}\ }\textbf {\bibinfo
  {volume} {2012}},\ \bibinfo {pages} {P03006} (\bibinfo {year}
  {2012})}\BibitemShut {NoStop}%
\bibitem [{\citenamefont {Chernyshev}\ and\ \citenamefont
  {Rozhkov}(2016)}]{PhysRevLett.116.017204}%
  \BibitemOpen
  \bibfield  {author} {\bibinfo {author} {\bibfnamefont {A.~L.}\ \bibnamefont
  {Chernyshev}}\ and\ \bibinfo {author} {\bibfnamefont {A.~V.}\ \bibnamefont
  {Rozhkov}},\ }\href {\doibase 10.1103/PhysRevLett.116.017204} {\bibfield
  {journal} {\bibinfo  {journal} {Phys. Rev. Lett.}\ }\textbf {\bibinfo
  {volume} {116}},\ \bibinfo {pages} {017204} (\bibinfo {year}
  {2016})}\BibitemShut {NoStop}%
\bibitem [{\citenamefont {Chernyshev}\ and\ \citenamefont
  {Brenig}(2015)}]{PhysRevB.92.054409}%
  \BibitemOpen
  \bibfield  {author} {\bibinfo {author} {\bibfnamefont {A.~L.}\ \bibnamefont
  {Chernyshev}}\ and\ \bibinfo {author} {\bibfnamefont {W.}~\bibnamefont
  {Brenig}},\ }\href {\doibase 10.1103/PhysRevB.92.054409} {\bibfield
  {journal} {\bibinfo  {journal} {Phys. Rev. B}\ }\textbf {\bibinfo {volume}
  {92}},\ \bibinfo {pages} {054409} (\bibinfo {year} {2015})}\BibitemShut
  {NoStop}%
\bibitem [{\citenamefont {Breuer}\ and\ \citenamefont
  {Petruccione}(2007)}]{breuer2007theory}%
  \BibitemOpen
  \bibfield  {author} {\bibinfo {author} {\bibfnamefont {H.}~\bibnamefont
  {Breuer}}\ and\ \bibinfo {author} {\bibfnamefont {F.}~\bibnamefont
  {Petruccione}},\ }\href {https://books.google.de/books?id=DkcJPwAACAAJ}
  {\emph {\bibinfo {title} {The Theory of Open Quantum Systems}}}\ (\bibinfo
  {publisher} {OUP Oxford},\ \bibinfo {year} {2007})\BibitemShut {NoStop}%
\bibitem [{\citenamefont {Lindblad}(1976)}]{lindblad1976}%
  \BibitemOpen
  \bibfield  {author} {\bibinfo {author} {\bibfnamefont {G.}~\bibnamefont
  {Lindblad}},\ }\href {https://projecteuclid.org:443/euclid.cmp/1103899849}
  {\bibfield  {journal} {\bibinfo  {journal} {Comm. Math. Phys.}\ }\textbf
  {\bibinfo {volume} {48}},\ \bibinfo {pages} {119} (\bibinfo {year}
  {1976})}\BibitemShut {NoStop}%
\bibitem [{\citenamefont {Altman}\ \emph {et~al.}(2003)\citenamefont {Altman},
  \citenamefont {Hofstetter}, \citenamefont {Demler},\ and\ \citenamefont
  {Lukin}}]{Altman_2003}%
  \BibitemOpen
  \bibfield  {author} {\bibinfo {author} {\bibfnamefont {E.}~\bibnamefont
  {Altman}}, \bibinfo {author} {\bibfnamefont {W.}~\bibnamefont {Hofstetter}},
  \bibinfo {author} {\bibfnamefont {E.}~\bibnamefont {Demler}}, \ and\ \bibinfo
  {author} {\bibfnamefont {M.~D.}\ \bibnamefont {Lukin}},\ }\href {\doibase
  10.1088/1367-2630/5/1/113} {\bibfield  {journal} {\bibinfo  {journal} {New
  Journal of Physics}\ }\textbf {\bibinfo {volume} {5}},\ \bibinfo {pages}
  {113} (\bibinfo {year} {2003})}\BibitemShut {NoStop}%
\bibitem [{\citenamefont {Chung}\ \emph {et~al.}(2021)\citenamefont {Chung},
  \citenamefont {de~Hond}, \citenamefont {Xiang}, \citenamefont
  {Cruz-Col\'on},\ and\ \citenamefont {Ketterle}}]{PhysRevLett.126.163203}%
  \BibitemOpen
  \bibfield  {author} {\bibinfo {author} {\bibfnamefont {W.~C.}\ \bibnamefont
  {Chung}}, \bibinfo {author} {\bibfnamefont {J.}~\bibnamefont {de~Hond}},
  \bibinfo {author} {\bibfnamefont {J.}~\bibnamefont {Xiang}}, \bibinfo
  {author} {\bibfnamefont {E.}~\bibnamefont {Cruz-Col\'on}}, \ and\ \bibinfo
  {author} {\bibfnamefont {W.}~\bibnamefont {Ketterle}},\ }\href {\doibase
  10.1103/PhysRevLett.126.163203} {\bibfield  {journal} {\bibinfo  {journal}
  {Phys. Rev. Lett.}\ }\textbf {\bibinfo {volume} {126}},\ \bibinfo {pages}
  {163203} (\bibinfo {year} {2021})}\BibitemShut {NoStop}%
\bibitem [{\citenamefont {de~Hond}\ \emph {et~al.}(2022)\citenamefont
  {de~Hond}, \citenamefont {Xiang}, \citenamefont {Chung}, \citenamefont
  {Cruz-Col\'on}, \citenamefont {Chen}, \citenamefont {Burton}, \citenamefont
  {Kennedy},\ and\ \citenamefont {Ketterle}}]{PhysRevLett.128.093401}%
  \BibitemOpen
  \bibfield  {author} {\bibinfo {author} {\bibfnamefont {J.}~\bibnamefont
  {de~Hond}}, \bibinfo {author} {\bibfnamefont {J.}~\bibnamefont {Xiang}},
  \bibinfo {author} {\bibfnamefont {W.~C.}\ \bibnamefont {Chung}}, \bibinfo
  {author} {\bibfnamefont {E.}~\bibnamefont {Cruz-Col\'on}}, \bibinfo {author}
  {\bibfnamefont {W.}~\bibnamefont {Chen}}, \bibinfo {author} {\bibfnamefont
  {W.~C.}\ \bibnamefont {Burton}}, \bibinfo {author} {\bibfnamefont {C.~J.}\
  \bibnamefont {Kennedy}}, \ and\ \bibinfo {author} {\bibfnamefont
  {W.}~\bibnamefont {Ketterle}},\ }\href {\doibase
  10.1103/PhysRevLett.128.093401} {\bibfield  {journal} {\bibinfo  {journal}
  {Phys. Rev. Lett.}\ }\textbf {\bibinfo {volume} {128}},\ \bibinfo {pages}
  {093401} (\bibinfo {year} {2022})}\BibitemShut {NoStop}%
\bibitem [{\citenamefont {Yarmohammadi}\ \emph
  {et~al.}(2024{\natexlab{b}})\citenamefont {Yarmohammadi}, \citenamefont
  {Sous}, \citenamefont {Bukov},\ and\ \citenamefont
  {Kolodrubetz}}]{yarmohammadi2024ultrafast}%
  \BibitemOpen
  \bibfield  {author} {\bibinfo {author} {\bibfnamefont {M.}~\bibnamefont
  {Yarmohammadi}}, \bibinfo {author} {\bibfnamefont {J.}~\bibnamefont {Sous}},
  \bibinfo {author} {\bibfnamefont {M.}~\bibnamefont {Bukov}}, \ and\ \bibinfo
  {author} {\bibfnamefont {M.~H.}\ \bibnamefont {Kolodrubetz}},\ }\href@noop {}
  {} (\bibinfo {year} {2024}{\natexlab{b}}),\ \Eprint
  {http://arxiv.org/abs/2402.12591} {arXiv:2402.12591 [cond-mat.str-el]}
  \BibitemShut {NoStop}%
\bibitem [{\citenamefont {Yarmohammadi}\ \emph {et~al.}(2021)\citenamefont
  {Yarmohammadi}, \citenamefont {Meyer}, \citenamefont {Fauseweh},
  \citenamefont {Normand},\ and\ \citenamefont
  {Uhrig}}]{yarmohammadi2020dynamical}%
  \BibitemOpen
  \bibfield  {author} {\bibinfo {author} {\bibfnamefont {M.}~\bibnamefont
  {Yarmohammadi}}, \bibinfo {author} {\bibfnamefont {C.}~\bibnamefont {Meyer}},
  \bibinfo {author} {\bibfnamefont {B.}~\bibnamefont {Fauseweh}}, \bibinfo
  {author} {\bibfnamefont {B.}~\bibnamefont {Normand}}, \ and\ \bibinfo
  {author} {\bibfnamefont {G.~S.}\ \bibnamefont {Uhrig}},\ }\href {\doibase
  10.1103/PhysRevB.103.045132} {\bibfield  {journal} {\bibinfo  {journal}
  {Phys. Rev. B}\ }\textbf {\bibinfo {volume} {103}},\ \bibinfo {pages}
  {045132} (\bibinfo {year} {2021})}\BibitemShut {NoStop}%
\bibitem [{\citenamefont {Yarmohammadi}\ \emph
  {et~al.}(2023{\natexlab{a}})\citenamefont {Yarmohammadi}, \citenamefont
  {Bukov},\ and\ \citenamefont {Kolodrubetz}}]{yarmohammadi2023nonequilibrium}%
  \BibitemOpen
  \bibfield  {author} {\bibinfo {author} {\bibfnamefont {M.}~\bibnamefont
  {Yarmohammadi}}, \bibinfo {author} {\bibfnamefont {M.}~\bibnamefont {Bukov}},
  \ and\ \bibinfo {author} {\bibfnamefont {M.~H.}\ \bibnamefont
  {Kolodrubetz}},\ }\href {\doibase 10.1103/PhysRevB.108.L140305} {\bibfield
  {journal} {\bibinfo  {journal} {Phys. Rev. B}\ }\textbf {\bibinfo {volume}
  {108}},\ \bibinfo {pages} {L140305} (\bibinfo {year}
  {2023}{\natexlab{a}})}\BibitemShut {NoStop}%
\bibitem [{\citenamefont {Yarmohammadi}\ \emph
  {et~al.}(2023{\natexlab{b}})\citenamefont {Yarmohammadi}, \citenamefont
  {Krebs}, \citenamefont {Uhrig},\ and\ \citenamefont
  {Normand}}]{PhysRevB.107.174415}%
  \BibitemOpen
  \bibfield  {author} {\bibinfo {author} {\bibfnamefont {M.}~\bibnamefont
  {Yarmohammadi}}, \bibinfo {author} {\bibfnamefont {M.}~\bibnamefont {Krebs}},
  \bibinfo {author} {\bibfnamefont {G.~S.}\ \bibnamefont {Uhrig}}, \ and\
  \bibinfo {author} {\bibfnamefont {B.}~\bibnamefont {Normand}},\ }\href
  {\doibase 10.1103/PhysRevB.107.174415} {\bibfield  {journal} {\bibinfo
  {journal} {Phys. Rev. B}\ }\textbf {\bibinfo {volume} {107}},\ \bibinfo
  {pages} {174415} (\bibinfo {year} {2023}{\natexlab{b}})}\BibitemShut
  {NoStop}%
\bibitem [{\citenamefont {Sentef}(2017)}]{PhysRevB.95.205111}%
  \BibitemOpen
  \bibfield  {author} {\bibinfo {author} {\bibfnamefont {M.~A.}\ \bibnamefont
  {Sentef}},\ }\href {\doibase 10.1103/PhysRevB.95.205111} {\bibfield
  {journal} {\bibinfo  {journal} {Phys. Rev. B}\ }\textbf {\bibinfo {volume}
  {95}},\ \bibinfo {pages} {205111} (\bibinfo {year} {2017})}\BibitemShut
  {NoStop}%
\bibitem [{\citenamefont {Grankin}\ \emph {et~al.}(2021)\citenamefont
  {Grankin}, \citenamefont {Hafezi},\ and\ \citenamefont
  {Galitski}}]{PhysRevB.104.L220503}%
  \BibitemOpen
  \bibfield  {author} {\bibinfo {author} {\bibfnamefont {A.}~\bibnamefont
  {Grankin}}, \bibinfo {author} {\bibfnamefont {M.}~\bibnamefont {Hafezi}}, \
  and\ \bibinfo {author} {\bibfnamefont {V.~M.}\ \bibnamefont {Galitski}},\
  }\href {\doibase 10.1103/PhysRevB.104.L220503} {\bibfield  {journal}
  {\bibinfo  {journal} {Phys. Rev. B}\ }\textbf {\bibinfo {volume} {104}},\
  \bibinfo {pages} {L220503} (\bibinfo {year} {2021})}\BibitemShut {NoStop}%
\bibitem [{\citenamefont {Kennes}\ \emph {et~al.}(2017)\citenamefont {Kennes},
  \citenamefont {Wilner}, \citenamefont {Reichman},\ and\ \citenamefont
  {Millis}}]{Kennes2017}%
  \BibitemOpen
  \bibfield  {author} {\bibinfo {author} {\bibfnamefont {D.~M.}\ \bibnamefont
  {Kennes}}, \bibinfo {author} {\bibfnamefont {E.~Y.}\ \bibnamefont {Wilner}},
  \bibinfo {author} {\bibfnamefont {D.~R.}\ \bibnamefont {Reichman}}, \ and\
  \bibinfo {author} {\bibfnamefont {A.~J.}\ \bibnamefont {Millis}},\ }\href
  {\doibase 10.1038/nphys4024} {\bibfield  {journal} {\bibinfo  {journal}
  {Nature Physics}\ }\textbf {\bibinfo {volume} {13}},\ \bibinfo {pages} {479}
  (\bibinfo {year} {2017})}\BibitemShut {NoStop}%
\bibitem [{\citenamefont {de~la Torre}\ \emph {et~al.}(2021)\citenamefont
  {de~la Torre}, \citenamefont {Kennes}, \citenamefont {Claassen},
  \citenamefont {Gerber}, \citenamefont {McIver},\ and\ \citenamefont
  {Sentef}}]{RevModPhys.93.041002}%
  \BibitemOpen
  \bibfield  {author} {\bibinfo {author} {\bibfnamefont {A.}~\bibnamefont
  {de~la Torre}}, \bibinfo {author} {\bibfnamefont {D.~M.}\ \bibnamefont
  {Kennes}}, \bibinfo {author} {\bibfnamefont {M.}~\bibnamefont {Claassen}},
  \bibinfo {author} {\bibfnamefont {S.}~\bibnamefont {Gerber}}, \bibinfo
  {author} {\bibfnamefont {J.~W.}\ \bibnamefont {McIver}}, \ and\ \bibinfo
  {author} {\bibfnamefont {M.~A.}\ \bibnamefont {Sentef}},\ }\href {\doibase
  10.1103/RevModPhys.93.041002} {\bibfield  {journal} {\bibinfo  {journal}
  {Rev. Mod. Phys.}\ }\textbf {\bibinfo {volume} {93}},\ \bibinfo {pages}
  {041002} (\bibinfo {year} {2021})}\BibitemShut {NoStop}%
\bibitem [{\citenamefont {Murakami}\ \emph {et~al.}(2023)\citenamefont
  {Murakami}, \citenamefont {Golež}, \citenamefont {Eckstein},\ and\
  \citenamefont {Werner}}]{murakami2023photoinduced}%
  \BibitemOpen
  \bibfield  {author} {\bibinfo {author} {\bibfnamefont {Y.}~\bibnamefont
  {Murakami}}, \bibinfo {author} {\bibfnamefont {D.}~\bibnamefont {Golež}},
  \bibinfo {author} {\bibfnamefont {M.}~\bibnamefont {Eckstein}}, \ and\
  \bibinfo {author} {\bibfnamefont {P.}~\bibnamefont {Werner}},\ }\href@noop {}
  {} (\bibinfo {year} {2023}),\ \Eprint {http://arxiv.org/abs/2310.05201}
  {arXiv:2310.05201 [cond-mat.str-el]} \BibitemShut {NoStop}%
\bibitem [{\citenamefont {Holstein}\ and\ \citenamefont
  {Primakoff}(1940)}]{PhysRev.58.1098}%
  \BibitemOpen
  \bibfield  {author} {\bibinfo {author} {\bibfnamefont {T.}~\bibnamefont
  {Holstein}}\ and\ \bibinfo {author} {\bibfnamefont {H.}~\bibnamefont
  {Primakoff}},\ }\href {\doibase 10.1103/PhysRev.58.1098} {\bibfield
  {journal} {\bibinfo  {journal} {Phys. Rev.}\ }\textbf {\bibinfo {volume}
  {58}},\ \bibinfo {pages} {1098} (\bibinfo {year} {1940})}\BibitemShut
  {NoStop}%
\bibitem [{Note1()}]{Note1}%
  \BibitemOpen
  \bibinfo {note} {The quantum fluctuations proportional to $1/\protect \sqrt
  {N}$ become negligible in a long chain $N =2001$, given that the phonon
  number scales with $N$}\BibitemShut {NoStop}%
\bibitem [{\citenamefont {Mori}\ \emph {et~al.}(2019)\citenamefont {Mori},
  \citenamefont {Marshall}, \citenamefont {Ahadi}, \citenamefont {Denlinger},
  \citenamefont {Stemmer},\ and\ \citenamefont {Lanzara}}]{Mori2019}%
  \BibitemOpen
  \bibfield  {author} {\bibinfo {author} {\bibfnamefont {R.}~\bibnamefont
  {Mori}}, \bibinfo {author} {\bibfnamefont {P.~B.}\ \bibnamefont {Marshall}},
  \bibinfo {author} {\bibfnamefont {K.}~\bibnamefont {Ahadi}}, \bibinfo
  {author} {\bibfnamefont {J.~D.}\ \bibnamefont {Denlinger}}, \bibinfo {author}
  {\bibfnamefont {S.}~\bibnamefont {Stemmer}}, \ and\ \bibinfo {author}
  {\bibfnamefont {A.}~\bibnamefont {Lanzara}},\ }\href {\doibase
  10.1038/s41467-019-13046-z} {\bibfield  {journal} {\bibinfo  {journal}
  {Nature Communications}\ }\textbf {\bibinfo {volume} {10}},\ \bibinfo {pages}
  {5534} (\bibinfo {year} {2019})}\BibitemShut {NoStop}%
\bibitem [{\citenamefont {Sen}\ and\ \citenamefont
  {Guo}(2020)}]{PhysRevMaterials.4.104802}%
  \BibitemOpen
  \bibfield  {author} {\bibinfo {author} {\bibfnamefont {S.}~\bibnamefont
  {Sen}}\ and\ \bibinfo {author} {\bibfnamefont {G.-Y.}\ \bibnamefont {Guo}},\
  }\href {\doibase 10.1103/PhysRevMaterials.4.104802} {\bibfield  {journal}
  {\bibinfo  {journal} {Phys. Rev. Mater.}\ }\textbf {\bibinfo {volume} {4}},\
  \bibinfo {pages} {104802} (\bibinfo {year} {2020})}\BibitemShut {NoStop}%
\bibitem [{\citenamefont {Singh}\ \emph {et~al.}(2019)\citenamefont {Singh},
  \citenamefont {Jouan}, \citenamefont {Herranz}, \citenamefont {Scigaj},
  \citenamefont {S{\'a}nchez}, \citenamefont {Benfatto}, \citenamefont
  {Caprara}, \citenamefont {Grilli}, \citenamefont {Saiz}, \citenamefont
  {Cou{\"e}do} \emph {et~al.}}]{Singh2019}%
  \BibitemOpen
  \bibfield  {author} {\bibinfo {author} {\bibfnamefont {G.}~\bibnamefont
  {Singh}}, \bibinfo {author} {\bibfnamefont {A.}~\bibnamefont {Jouan}},
  \bibinfo {author} {\bibfnamefont {G.}~\bibnamefont {Herranz}}, \bibinfo
  {author} {\bibfnamefont {M.}~\bibnamefont {Scigaj}}, \bibinfo {author}
  {\bibfnamefont {F.}~\bibnamefont {S{\'a}nchez}}, \bibinfo {author}
  {\bibfnamefont {L.}~\bibnamefont {Benfatto}}, \bibinfo {author}
  {\bibfnamefont {S.}~\bibnamefont {Caprara}}, \bibinfo {author} {\bibfnamefont
  {M.}~\bibnamefont {Grilli}}, \bibinfo {author} {\bibfnamefont
  {G.}~\bibnamefont {Saiz}}, \bibinfo {author} {\bibfnamefont {F.}~\bibnamefont
  {Cou{\"e}do}},  \emph {et~al.},\ }\href {\doibase 10.1038/s41563-019-0354-z}
  {\bibfield  {journal} {\bibinfo  {journal} {Nature Materials}\ }\textbf
  {\bibinfo {volume} {18}},\ \bibinfo {pages} {948} (\bibinfo {year}
  {2019})}\BibitemShut {NoStop}%
\bibitem [{\citenamefont {Lindemann}(1910)}]{Lindemann1910}%
  \BibitemOpen
  \bibfield  {author} {\bibinfo {author} {\bibfnamefont {F.~A.}\ \bibnamefont
  {Lindemann}},\ }\href {https://ntrs.nasa.gov/citations/19840027015}
  {\bibfield  {journal} {\bibinfo  {journal} {Phys. Z.}\ }\textbf {\bibinfo
  {volume} {11}},\ \bibinfo {pages} {609} (\bibinfo {year} {1910})}\BibitemShut
  {NoStop}%
\bibitem [{\citenamefont {Kimel}\ \emph {et~al.}(2022)\citenamefont {Kimel},
  \citenamefont {Zvezdin}, \citenamefont {Sharma}, \citenamefont {Shallcross},
  \citenamefont {de~Sousa}, \citenamefont {García-Martín}, \citenamefont
  {Salvan}, \citenamefont {Hamrle}, \citenamefont {Stejskal}, \citenamefont
  {McCord} \emph {et~al.}}]{Kimel_2022}%
  \BibitemOpen
  \bibfield  {author} {\bibinfo {author} {\bibfnamefont {A.}~\bibnamefont
  {Kimel}}, \bibinfo {author} {\bibfnamefont {A.}~\bibnamefont {Zvezdin}},
  \bibinfo {author} {\bibfnamefont {S.}~\bibnamefont {Sharma}}, \bibinfo
  {author} {\bibfnamefont {S.}~\bibnamefont {Shallcross}}, \bibinfo {author}
  {\bibfnamefont {N.}~\bibnamefont {de~Sousa}}, \bibinfo {author}
  {\bibfnamefont {A.}~\bibnamefont {García-Martín}}, \bibinfo {author}
  {\bibfnamefont {G.}~\bibnamefont {Salvan}}, \bibinfo {author} {\bibfnamefont
  {J.}~\bibnamefont {Hamrle}}, \bibinfo {author} {\bibfnamefont
  {O.}~\bibnamefont {Stejskal}}, \bibinfo {author} {\bibfnamefont
  {J.}~\bibnamefont {McCord}},  \emph {et~al.},\ }\href {\doibase
  10.1088/1361-6463/ac8da0} {\bibfield  {journal} {\bibinfo  {journal} {Journal
  of Physics D: Applied Physics}\ }\textbf {\bibinfo {volume} {55}},\ \bibinfo
  {pages} {463003} (\bibinfo {year} {2022})}\BibitemShut {NoStop}%
\bibitem [{\citenamefont {Hennecke}\ \emph {et~al.}(2022)\citenamefont
  {Hennecke}, \citenamefont {Schick}, \citenamefont {Sidiropoulos},
  \citenamefont {Willems}, \citenamefont {Heilmann}, \citenamefont {Bock},
  \citenamefont {Ehrentraut}, \citenamefont {Engel}, \citenamefont {Hessing},
  \citenamefont {Pfau} \emph {et~al.}}]{PhysRevResearch.4.L022062}%
  \BibitemOpen
  \bibfield  {author} {\bibinfo {author} {\bibfnamefont {M.}~\bibnamefont
  {Hennecke}}, \bibinfo {author} {\bibfnamefont {D.}~\bibnamefont {Schick}},
  \bibinfo {author} {\bibfnamefont {T.}~\bibnamefont {Sidiropoulos}}, \bibinfo
  {author} {\bibfnamefont {F.}~\bibnamefont {Willems}}, \bibinfo {author}
  {\bibfnamefont {A.}~\bibnamefont {Heilmann}}, \bibinfo {author}
  {\bibfnamefont {M.}~\bibnamefont {Bock}}, \bibinfo {author} {\bibfnamefont
  {L.}~\bibnamefont {Ehrentraut}}, \bibinfo {author} {\bibfnamefont
  {D.}~\bibnamefont {Engel}}, \bibinfo {author} {\bibfnamefont
  {P.}~\bibnamefont {Hessing}}, \bibinfo {author} {\bibfnamefont
  {B.}~\bibnamefont {Pfau}},  \emph {et~al.},\ }\href {\doibase
  10.1103/PhysRevResearch.4.L022062} {\bibfield  {journal} {\bibinfo  {journal}
  {Phys. Rev. Res.}\ }\textbf {\bibinfo {volume} {4}},\ \bibinfo {pages}
  {L022062} (\bibinfo {year} {2022})}\BibitemShut {NoStop}%
\bibitem [{\citenamefont {Gray}\ \emph {et~al.}(2024)\citenamefont {Gray},
  \citenamefont {Deng}, \citenamefont {Tian}, \citenamefont {Chilcote},
  \citenamefont {Brahlek},\ and\ \citenamefont {Wu}}]{gray2024timeresolved}%
  \BibitemOpen
  \bibfield  {author} {\bibinfo {author} {\bibfnamefont {I.}~\bibnamefont
  {Gray}}, \bibinfo {author} {\bibfnamefont {Q.}~\bibnamefont {Deng}}, \bibinfo
  {author} {\bibfnamefont {Q.}~\bibnamefont {Tian}}, \bibinfo {author}
  {\bibfnamefont {M.}~\bibnamefont {Chilcote}}, \bibinfo {author}
  {\bibfnamefont {M.}~\bibnamefont {Brahlek}}, \ and\ \bibinfo {author}
  {\bibfnamefont {L.}~\bibnamefont {Wu}},\ }\href@noop {} {} (\bibinfo {year}
  {2024}),\ \Eprint {http://arxiv.org/abs/2404.05020} {arXiv:2404.05020
  [cond-mat.mtrl-sci]} \BibitemShut {NoStop}%
\bibitem [{\citenamefont {Puviani}\ \emph {et~al.}(2023)\citenamefont
  {Puviani}, \citenamefont {Haenel},\ and\ \citenamefont
  {Manske}}]{PhysRevB.107.094501}%
  \BibitemOpen
  \bibfield  {author} {\bibinfo {author} {\bibfnamefont {M.}~\bibnamefont
  {Puviani}}, \bibinfo {author} {\bibfnamefont {R.}~\bibnamefont {Haenel}}, \
  and\ \bibinfo {author} {\bibfnamefont {D.}~\bibnamefont {Manske}},\ }\href
  {\doibase 10.1103/PhysRevB.107.094501} {\bibfield  {journal} {\bibinfo
  {journal} {Phys. Rev. B}\ }\textbf {\bibinfo {volume} {107}},\ \bibinfo
  {pages} {094501} (\bibinfo {year} {2023})}\BibitemShut {NoStop}%
\bibitem [{\citenamefont {Chen}\ \emph {et~al.}(2021)\citenamefont {Chen},
  \citenamefont {Dong},\ and\ \citenamefont
  {Qiu}}]{https://doi.org/10.1002/qute.202100052}%
  \BibitemOpen
  \bibfield  {author} {\bibinfo {author} {\bibfnamefont {Z.}~\bibnamefont
  {Chen}}, \bibinfo {author} {\bibfnamefont {G.}~\bibnamefont {Dong}}, \ and\
  \bibinfo {author} {\bibfnamefont {J.}~\bibnamefont {Qiu}},\ }\href {\doibase
  https://doi.org/10.1002/qute.202100052} {\bibfield  {journal} {\bibinfo
  {journal} {Advanced Quantum Technologies}\ }\textbf {\bibinfo {volume} {4}},\
  \bibinfo {pages} {2100052} (\bibinfo {year} {2021})}\BibitemShut {NoStop}%
\end{thebibliography}%
\end{document}